\documentstyle[12pt,eepic,epic]{article}
\newlength{\minitwocolumn}
\def\l{\ell}
\def\pv{\hat{P}_V}
\def\fv{\hat{f}_V}
\def\q{\hat{q}}
\def\s{\hat{s}}
\def\u{\hat{u}} 
\def\ml{\hat{m}_\ell}
\def\ms{\hat{m}_s}
\def\md{\hat{m}_d}
\def\mc{\hat{m}_c}
\def\muq{\hat{m}_u}
\def\mq{\hat{m}_q}
\def\mv{\hat{m}_V}
\def\mui{\hat{m}_{u^i}}

\title{Determination of the CKM unitarity triangle by 
	$B \rightarrow X_d \, \l^+ \, \l^- $ decay}
\author{\vspace{1cm}\\
        {\bf L. T. Handoko} \thanks{On leave from P3FT-LIPI, Indonesia. 
        E-mail address : handoko@theo.phys.sci.hiroshima-u.ac.jp} \\
        Department of Physics, Hiroshima University \\
        1-3-1 Kagamiyama, Higashi Hiroshima - 739, Japan\\
        \vspace{3mm}\\}
\date{}

\begin{document}

\setlength{\baselineskip}{24pt}

\maketitle
\begin{picture}(0,0)
       \put(325,260){HUPD-9713}
       \put(325,245){June 1997}
\end{picture}
\vspace{-24pt}

\setlength{\baselineskip}{8mm}
\thispagestyle{empty}

\begin{abstract}

I examine a possibility to extract the angle $\gamma$ of the
Cabibbo-Kobayashi-Maskawa unitarity triangle by the inclusive  
$B \rightarrow X_d \, \l^+ \, \l^- $ decay. 
An independent information for the angle is expected 
from the non-trivial contribution induced in $u \bar{u}$ and
$c \bar{c}$ loops.
The contributions induce CP asymmetry which has high 
sensitivity to the angle $\gamma$ in the whole dilepton invariant 
mass region. Particularly, in the low dilepton invariant mass region,  
the sensitivity is also recognized in the branching-ratio 
with any dilepton final states and the lepton polarization
asymmetry with dimuon final state.
In the high dilepton invariant mass region, the sensitivity 
is tiny in all measurements with the exception of the CP asymmetry, 
that make them to be good probes to 
confirm the measurement of ${V_{td}}^\ast V_{tb}$ in addition to 
the present data from $B_d^0-\bar{B}_d^0$ mixing.
The decay rate and asymmetries are examined  in the Standard Model 
with taking into account the long-distance contributions due
to vector-mesons as well as its momentum dependences that would 
reduce the long-distance backgrounds in the channel.
\end{abstract}

\clearpage

\section{\bf Introduction}
\label{sec:introduction}

The hope that $B \rightarrow X_s \, \l^+ \, \l^-$ decay will be within 
experimental reach in the near future \cite{bphysic} encourage me to
consider $B \rightarrow X_d \, \l^+ \, \l^-$ decay. Both decays are 
important probes of the effective Hamiltonian 
governing the flavor-changing neutral current (FCNC) transition 
$b \rightarrow q \, \l^+ \, \l^-$ ($q = s, d$) in the Standard Model (SM)
as written below \cite{bqll} 
\begin{eqnarray}
	{\cal H}_{\rm eff} & = & \frac{G_F \, \alpha}{\sqrt{2} \, \pi} \, 
		V_{tq}^\ast \, V_{tb} \, \left\{ 
		{C_9}^{\rm eff} \, 
			\left[ \bar{q} \, \gamma_\mu \, L \, b \right] \, 
			\left[ \bar{\l} \, \gamma^\mu \, \l \right]
		+ C_{10} \,  \left[ \bar{q} \, \gamma_\mu \, L \, b \right] \, 
			\left[ \bar{\l} \, \gamma^\mu \, \gamma_5 \, \l \right]
		\right. \nonumber \\
	& &	\; \; \; \; \; \; \; \; \; \; \; \; \; \; \; \; \; \; \; \; \; \; \; \; \; \; \left. 
		- 2 \, {C_7}^{\rm eff} \, 
			\left[ \bar{q} \, i \, \sigma_{\mu \nu} \, 
			\frac{\q^\nu}{\s} \left( R + \mq \, L \right) \, b
			\right]	
			\left[ \bar{\l} \, \gamma^\mu \, \l \right]
		\right\} \; .
		\label{eq:heff}
\end{eqnarray}
where $L/R \equiv {(1 \mp \gamma_5)}/2$, $q^\mu$ denotes four-momentum 
of the dilepton, $s = q^2$. Notation with hat on the top means it is 
normalized with $m_b$. 

\begin{figure}[t]
	\unitlength 1mm
	\begin{center}
	\begin{picture}(60,40)
		\put(5,5){\vector(0,1){50}}
		\put(5,5){\vector(1,0){75}}
		\thicklines
		\put(65,5.1){\line(-1,0){60}}
		\put(5,5){\line(1,2){20}}
		\put(25,45){\line(1,-1){40}}
		\multiput(25,5)(0,5){8}{\line(0,1){1.8}}
		\multiput(5,45)(5,0){4}{\line(1,0){1.8}}
		\put(5,2){\makebox(0,0){$(0,0)$}}
		\put(65,2){\makebox(0,0){$(1,0)$}}
		\put(25,47){\makebox(0,0){$(\rho,\eta)$}}
		\put(25,38){\makebox(0,0){$\alpha$}}
		\put(57,8){\makebox(0,0){$\beta$}}
		\put(10,8){\makebox(0,0){$\gamma$}}
		\put(25,2){\makebox(0,0){$(\rho,0)$}}
		\put(-2,45){\makebox(0,0){$(0,\eta)$}}
		\put(12,27){\makebox(0,0){$r$}}
		\put(50,27){\makebox(0,0){$x$}}
	\end{picture}
	\caption{The CKM unitarity triangle on the $\rho-\eta$ plane.}
	\label{fig:ckm}
	\end{center}
\end{figure}
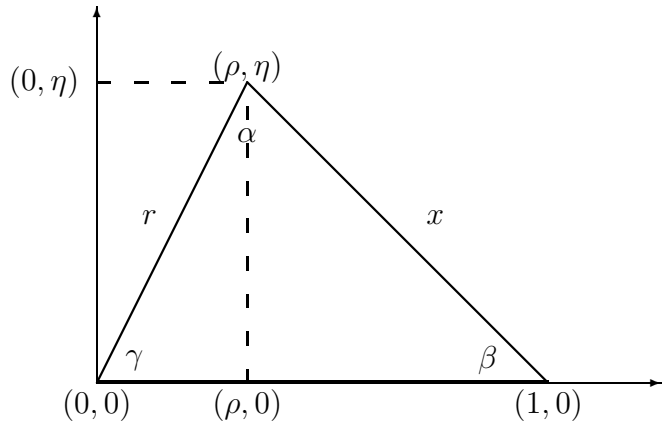

Theoretically, the most important interest in the 
$b \rightarrow d \, \l^+ \, \l^-$ decay is, 
the matrix element contains un-negligible terms induced by 
continuum part of $u\bar{u}$ and $c\bar{c}$ loops  
proportional to $V_{ud}^\ast \, V_{ub}$ and 
$V_{cd}^\ast \, V_{cb}$ \cite{cpvf}. These terms should give 
non-trivial contributions in the Wilson coefficient ${C_9}^{\rm eff}$ 
which induce CP violation in the channel. 
I call the contribution as CP violation factor (${C_9}^{\rm CP}$) 
throughout this paper. I will show that it can be utilized
to determine the length $x$ and the angle $\gamma$ of the 
Cabibbo-Kobayashi-Maskawa (CKM) unitarity triangle in Fig. \ref{fig:ckm}
at once. In this meaning, for example, the radiative 
$B \rightarrow X_d \, \gamma$ decay is not so useful 
since it gives only information for the length $x$ that have already been 
measured well in the $B_d^0-\bar{B}_d^0$ mixing.
On the other hand, in the $b \rightarrow s \, \l^+ \, \l^-$ decay, 
${C_9}^{\rm CP}$ is strongly suppressed due to the GIM mechanism.
Generally, rare $B$ decays are clean processes to extract the CKM 
matrix elements, because non-perturbative effects in the decays 
are possibly tiny, less than few percents as shown in \cite{hqet} by using 
heavy-quark effective theory approach \footnote{Remark that the approach 
is reliable only in the low dilepton mass region, but it is sufficient 
to justify the statement since in the high dilepton mass region 
the perturbative calculation is good.}.
In the analysis, I also utilize the experiment result of $x_d$
in the $B_d^0-\bar{B}_d^0$ mixing.

The purpose of this paper is to show a possibility to give an independent 
measurement for the angle $\gamma$ of CKM unitarity triangle in the SM 
by observing the channel. The calculation is done with taking into 
account the $q-$dependence in the long-distance (LD) contributions 
due to the vector mesons \cite{lde}. It is well known that including 
the $q-$dependence, which has not been considered in the previous 
papers, will reduce the background due to LD contributions \cite{ldeq}.

This paper is organized as follows. First I briefly describe the 
non-trivial contributions due to resonance and continuum parts 
of $u\bar{u}$ and $c\bar{c}$ loops which induce ${C_9}^{\rm CP}$. 
Next I consider the phenomenology of the contributions and its 
relation with the CKM unitarity triangle. Before going to summary, 
I analyse the decay rate and asymmetries, i.e. forward-backward (FB) asymmetry 
($\bar{\cal A}_{\rm FB}$), CP asymmetry ($\bar{\cal A}_{\rm CP}$) 
and lepton-polarization (LP) asymmetry ($\bar{\cal A}_{\rm LP}$) 
in the channel.

\section{\bf CP violation factor}
\label{sec:cpvf}

The effective Hamiltonian in Eq. (\ref{eq:heff}) describes both inclusive 
$b \rightarrow q \, \l^+ \, \l^-$ decays by replacing $q$ with $s$ or $d$
quark respectively. In the SM, the QCD corrected Wilson 
coefficients enter in the physical decay amplitude above have been 
calculated up to next-to leading order (NLO) for ${C_9}^{\rm eff}$ and 
leading order (LO) for ${C_7}^{\rm eff}$ \cite{qcd}, while $C_{10}$ does 
not receive any correction at all. Some corrections due to continuum 
parts of $u\bar{u}$ and $c\bar{c}$ continuums and the resonances of 
the vector-mesons will enter only in the coefficient ${C_9}^{\rm eff}$. 
Remind that the contribution of $u\bar{u}$ loop in ${C_7}^{\rm eff}$ is 
suppressed due to the GIM mechanism \cite{inamilim}.

Before going to give ${C_9}^{\rm eff}$, related with 
$u\bar{u}$ and $c\bar{c}$ loops, let me mention the operators govern 
the $b \rightarrow q \, u^i \, \bar{u}^i$ processes, 
\begin{eqnarray}
	{\cal O}_1 & = & 
		\left( \bar{q}_\alpha \, \gamma_\mu \, L \, b_\alpha 
			\right) \, 
		\sum_{i=1,2} \, 
		\left( \bar{u}_\beta^i \, \gamma^\mu \, L \, u_\beta^i 
			\right) \; , 
		\label{eq:o1} \\
	{\cal O}_2 & = & 
		\left( \bar{q}_\alpha \, \gamma_\mu \, L \, b_\beta
			\right) \, 
		\sum_{i=1,2} \, 
		\left( \bar{u}_\beta^i \, \gamma^\mu \, L \, u_\alpha^i 
			\right) \; , 
		\label{eq:o2}
\end{eqnarray}
after doing Fierz transformation. Here, $u^1 = u$, $u^2 = c$ and the lower
suffixes denote the color.
For $q = s$, $u\bar{u}$ loop contribution can be ignored \cite{bqll, qcd}, 
while for $q = d$ the situation is quite different. The reason is, 
both operators above are proportional to the CKM matrix element 
${V_{u^iq}^\ast \, V_{u^ib}}/{V_{tq}^\ast \, V_{tb}}$ if we normalize the 
amplitude with $V_{tq}^\ast \, V_{tb}$ as usual, then 
\begin{equation}
	\left| \frac{V_{u^is}^\ast \, V_{u^ib}}{V_{ts}^\ast \, V_{tb}} \right| 
		\sim \left\{ 
		\begin{array}{lcl}
			O(\lambda^2) 	& , & u^i = u \\
			O(1) 		& , & u^i = c
		\end{array}
		\right. \; , 
	\label{eq:ratiockms}
\end{equation} 
for the former channel, and
\begin{equation}
	\left| \frac{V_{u^id}^\ast \, V_{u^ib}}{V_{td}^\ast \, V_{tb}} \right| 
		\sim \left\{ 
		\begin{array}{lcl}
			O(1) 		& , & u^i = u \\
			O(1) 		& , & u^i = c
		\end{array}
		\right. \; , 
	\label{eq:ratiockmd}
\end{equation} 
for the later one. $\lambda$ is a parameter in Wolfenstein parametrization of 
CKM matrix \cite{wolfenstein} and the world average is
$\lambda \sim 0.22$ \cite{pdg}.

Involving the continuum as well as resonance parts of $u\bar{u}$ and 
$c\bar{c}$ loops and NLO QCD correction into the calculation gives 
\begin{equation}
	{C_9}^{\rm eff} = {C_9}^{\rm NLO} \, \left[ 
		1 + \frac{\alpha_s(\mu)}{\pi} \omega(\s) \right]
		+ {C_9}^{\rm con}(\s) + {C_9}^{\rm res}(\s) \; ,
	\label{eq:c9eff}
\end{equation}
where 
\begin{eqnarray}
	{C_9}^{\rm con}(\s) & = & 
		\left[ \left( 1 + 
		\frac{V_{uq}^\ast \, V_{ub}}{V_{tq}^\ast \, V_{tb}} \right) 
		g(\mc, \s) 
		- \frac{V_{uq}^\ast \, V_{ub}}{V_{tq}^\ast \, V_{tb}} 
			g(\muq, \s) \right] 
	\nonumber \\
	& & \; \; \; \; \; \times \left(3 \, C_1 + C_2 + 3 \, C_3
                	+ C_4 + 3 \, C_5 + C_6 \right)
		\nonumber \\
        & & - \frac{1}{2} g(1,\s) \left( 4 \, C_3 + 4 \, C_4 + 3 \,
                C_5 + C_6 \right) 
		\nonumber \\
        & & - \frac{1}{2} g(0,\s) \left( C_3 + 3 \, C_4 \right) 
        	+ \frac{2}{9} \left( 3 \, C_3 + C_4 +
                	3 \, C_5 + C_6 \right) \; , 
	\label{eq:c9con} \\
	{C_9}^{\rm res}(\s) & = & -\frac{16 \, \pi^2}{9} 
		\left( 3 \, C_1 + C_2 + 3 \, C_3 + C_4 + 
			3 \, C_5 + C_6 \right)
	\nonumber \\	
	& &	\times \left[ 
		\left( 1 + 
		\frac{V_{uq}^\ast \, V_{ub}}{V_{tq}^\ast \, V_{tb}} \right) 
		\sum_{V=\psi,\cdots} F_V(\s) 
		- \frac{V_{uq}^\ast \, V_{ub}}{V_{tq}^\ast \, V_{tb}}
		\sum_{V=\rho,\omega} F_V(\s) \right] \; . 
	\label{eq:c9res}
\end{eqnarray}
The readers should refer \cite{qcd} for ${C_9}^{\rm NLO}$ and 
\begin{eqnarray}
	\omega(\s) & = & -\frac{2}{9} \pi^2 
		- \frac{4}{3} {\rm Li}_2(\s) 
		- \frac{2}{3} \ln \s \, \ln(1 - \s) 
		- \frac{5 + 4 \, \s}{3 ( 1 + 2 \, \s)} \, 
			\ln(1 - \s) 
	\nonumber \\
	& & -\frac{2 \, \s \, (1 + \s)(1 - 2 \, \s)}{
		3 (1 - \s)^2 (1 + 2 \, \s)} \ln \s 
		+ \frac{5 + 9 \, \s - 6 \, \s^2}{6 (1 - \s)(1 + 2 \s)} 
		\, , 
	\label{eq:omega}
\end{eqnarray}
represents the $O(\alpha_s)$ correction from the one gluon 
exchange in the matrix element of ${\cal O}_9$.
The function $g(\mui, \s)$ which describes the continuum part
of $u^i\bar{u}^i$ pair contribution is 
\begin{eqnarray}
	g(\mui, \s) & = & -\frac{8}{9} \, \ln \left( \frac{m_b}{\mu} \right) 
		- \frac{8}{9} \, \ln (\mui) + \frac{8}{27} 
		+ \frac{16}{9} \, \frac{\mui^2}{\s}  
		- \frac{2}{9} \, \left( 2 + 
	\frac{4 \mui^2}{\s} \right) \sqrt{ \left| 1 - \frac{4 \mui^2}{\s} \right| } 
	\nonumber \\
	& & \times \left[ \Theta \left( 1 - \frac{4 \mui^2}{\s} \right) 
		\left( \ln \frac{1 + \sqrt{ 1 - {4 \mui^2}/{\s}}}{
			1 - \sqrt{ 1 - {4 \mui^2}/{\s}}} - i \, \pi \right)
		\right. 
	\nonumber \\
	& & \left. \; \; \; \; \; \; 
		+  \Theta \left( \frac{4 \mui^2}{\s} - 1 \right) \, 2 \, 
		\arctan \frac{1}{\sqrt{ {4 \mui^2}/{\s} - 1}}	
		\right] \; , 
	\label{eq:g} \\
	g(0, \s) & = & \frac{8}{27} 
		- \frac{8}{9} \ln \left( \frac{m_b}{\mu} \right)
		- \frac{4}{9} \ln \s 
		+ \frac{4}{9} \, i \, \pi \; .
	\label{eq:g0}
\end{eqnarray}

In the resonance part ${C_9}^{\rm res}$, I put the relative phase 
to be zero because of unitarity constraint in the Argand plot of
the transition amplitude \cite{lde}. $F_V(\s)$ is the Breit-Wigner 
resonance form 
\begin{equation}
	F_V(\s) = \frac{{\fv^2(\s)}/\s}{
		\s - \mv^2 + i \, \mv \, \hat{\Gamma}_V} \, .
	\label{eq:fs}
\end{equation}
$\fv(\s)$ describes the momentum dependence of coupling strength of 
vector interaction in $\gamma-V$ transition, i.e. 
\begin{equation}
	\left< 0 \left| \bar{u}^i \, \gamma_\mu \, u^i \right| V(q) 
		\right> \equiv f_V(q^2) \, \epsilon_\mu \; ,
	\label{eq:mefv}	
\end{equation}
and has been derived as follows \cite{ldeq}
\begin{equation}
	\frac{\fv(\s)}{\fv(0)} = 1 + 
		\frac{\s}{\pv} \left[ P^\prime_V - 
		P^{\prime \prime}_V(\s) \right] \; ,
	\label{eq:fvq} 
\end{equation}
under an assumption that the vector-mesons are bound-states of the 
pair $u^i \bar{u}^i$. Here, 
\begin{equation}
	P^{\prime \prime}_V(\s) = \frac{\mui^2}{4 \, \pi^2 \, \s} \left[
		-4 - \frac{5 \, \s}{3 \, \mui^2} 
		+ 4 \left( 1 + \frac{\s}{2 \, \mui^2} \right) 
		\sqrt{\frac{4 \, \mui^2}{\s} - 1} 
		\arctan \frac{1}{\displaystyle 
		\sqrt{{4 \, \mui^2}/{\s} - 1}}
	\right] \; , 
	\label{eq:pvq}
\end{equation}
is obtained from a dispersion relation involving the 
imaginary part of quark-loop diagram, while $P_V$ and $P^\prime_V$ 
are the subtraction constants. 
Kinematically the above interpolation equation of $\hat{f}_V$ is valid only 
for $0 \leq \s \leq \mv^2$ region. For $\s > \mv^2$ region I 
take same assumption with \cite{ldeq}, that is $\fv(\s>\mv^2) = \fv(\mv^2)$.
In principle, the ratio in Eq. (\ref{eq:fvq}) should be obtained from 
the known data of $V$ production cross-section by off-shell and on-shell photons.

\begin{table}[t]
        \begin{center}
        \begin{tabular}{lp{15mm}p{15mm}p{15mm}p{15mm}p{15mm}p{15mm}}
        \hline \hline
        \multicolumn{1}{c}{\raisebox{-1.5ex}{Parameter}} & 
                \multicolumn{6}{c}{Vector-meson} \\
		\cline{2-7} 
		\multicolumn{1}{c}{} & 
		\multicolumn{1}{c}{$\rho$} & 
		\multicolumn{1}{c}{$\omega$} & 
		\multicolumn{1}{c}{$\psi$} & 
		\multicolumn{1}{c}{$\psi^\prime$} & 
		\multicolumn{1}{c}{$\psi^{\prime \prime}$}& 
		\multicolumn{1}{c}{$\psi^{\prime \prime \prime}$}\\ 
        \hline
        $M_V$ (MeV)	& 768.5
			& 781.94
			& 3096.88 
			& 3686.00 
			& 3769.9	
			& 4040
			\\
        $\Gamma_{e^+ e^-}$ (keV) & 6.77
			& 0.6
			& 5.26
			& 2.14
	   		& 0.26   
			& 0.75
			\\
        $\Gamma_V$ (MeV)& 150.7
			& 8.43
			& 0.087
			& 0.277
			& 23.6
			& 52
		      	\\
	$P^{\prime \prime}_V(\mv^2)$ & -0.00243
			& -0.00243
			& -0.02734
			& -0.01463
			& -0.01374
			& -0.01153
			\\
        \hline 
	$\pv$		& 0.01339 
			& 0.01387 
			& 0.01577
			& 0.01830 
			& 0.01885
			& 0.02080 
			\\
	$\fv(0)$	& 0.00662
			& 0.00202
			& 0.01795
			& 0.01487
			& 0.00536
			& 0.01010
			\\
	\hline \hline
        \end{tabular}
        \caption{The experimental (central) values for each vector-meson under 
	consideration (upper table) and the determined constants 
	under these values (lower table).}
        \label{tab:vector}
        \end{center}
\end{table}

The subtraction constants in Eq. (\ref{eq:fvq}) are written in 
the lower table of Tab. (\ref{tab:vector}) for each vector meson. 
The results are determined by using the data in the upper table, 
putting $m_V \sim (2 \, m_{u^i})$ and $P^\prime_V = 0.043$ 
for all $V$'s. Unfortunately, there is no data of photoproduction 
for higher excited states of $\psi$, so let me use same average value 
$|{\fv(0)}/{\fv(\mv^2)}| \sim 0.35$ for 
$V = \psi, \psi^\prime, \psi^{\prime \prime}, 
\psi^{\prime \prime \prime}$ and 
$|{\fv(0)}/{\fv(\mv^2)}| \sim 0.92$ for $V = \rho, \omega$ 
which fit the data on $\rho, \omega$ and $\psi$ \cite{ldeq}. 
This fact is also the reason why other resonances higher 
than $\psi^{\prime \prime \prime}$ are not considered here. 
Otherwise, $\fv(\mv^2)$ can be obtained from the data 
on leptonic width \cite{pdg}, that is 
\begin{equation}
	\fv^2(\mv^2) = \frac{27 \, \mv^3}{16 \, \pi \, \alpha^2}
	\hat{\Gamma}(V \rightarrow \l^+ \, \l^-) \; ,
	\label{eq:vll}
\end{equation}
then $\fv(0)$ would follow respectively as written in 
Tab. \ref{tab:vector}.

From Eqs. (\ref{eq:c9con}) and (\ref{eq:c9res}), it is obvious
that ${V_{uq}^\ast \, V_{ub}}/{V_{tq}^\ast \, V_{tb}}$ would
induce CP violation in the channel. For convenience, these terms 
can be collected as $\left( {V_{uq}^\ast \, V_{ub}}/{V_{tq}^\ast \, V_{tb}} 
\right) {C_9}^{\rm CP}(\s)$, with 
\begin{eqnarray}
	{C_9}^{\rm CP}(\s) & = &
		\left( 3 \, C_1 + C_2 + 3 \, C_3
                	+ C_4 + 3 \, C_5 + C_6 \right)
	\nonumber \\
	& & \times \left[ g(\mc, \s) - g(\muq, \s) 
		- \frac{16 \, \pi^2}{9} \left(
		\sum_{V=\psi,\cdots} F_V(\s) - 
		\sum_{V=\rho,\omega} F_V(\s) \right) \right]\; .
	\label{eq:fcp}	
\end{eqnarray}

\begin{table}[t]
        \begin{center}
        \begin{tabular}{p{6cm}l}
        \hline \hline
        \multicolumn{1}{c}{Parameter} & 
                \multicolumn{1}{c}{Value}     \\
        \hline 
        $m_W$                   & $80.26 \pm 0.16$ (GeV)        \\
        $m_Z$                   & $91.19 \pm 0.002$ (GeV)       \\
	$m_u$			& $0.005$ (GeV) \\
	$m_d$			& $0.139$ (GeV) \\
        $m_c$                   & $1.4$ (GeV) \\
        $m_b$                   & $4.8$ (GeV) \\
        $m_t$                   & $175 \pm 9$ (GeV)     \\
	$m_e$			& $0.511$ (MeV) \\
	$m_\mu$			& $105.66$ (MeV) \\
	$m_\tau$		& $1777 ^{+0.30}_{-0.27} $ (MeV) \\	 
        $\mu$                   & $5^{+5.0}_{-2.5}$ (GeV)       \\
        $\Lambda_{QCD}^{(5)}$   & $0.214^{+0.066}_{-0.054}$ (GeV)        \\
        $\alpha_{QED}^{-1}$     & 129           \\
        $\alpha_s (m_Z) $       & $0.117 \pm 0.005$ \\
        $\sin^2 \theta_w $      & $0.2325$ \\
	$m_{B^0}$ 		& $5279.2 \pm 1.8$ (MeV) \\ 
	$\tau_{B^0}$		& $1.28 \pm 0.06$ (ps) \\
	$\eta_{\rm QCD}$	& $0.55$ \\
	$\sqrt{{f_{B_d}}^2 \, B_{B_d}}$ & $173 \pm 40$ (MeV) \\
        ${\cal B} (B \rightarrow X_c \, \l \, \bar{\nu}) $ & 
		$(10.4 \pm 0.4) \%$ \\
	$x_d $			& $0.73 \pm 0.05$ \\
        \hline \hline
        \end{tabular}
        \caption{The values of parameters used throughout the paper.}
        \label{tab:parameters}
        \end{center}
\end{table}

\begin{figure}[t]
        \unitlength 1mm
        \begin{center}
\setlength{\unitlength}{0.240900pt}
\begin{picture}(1500,900)(0,0)
\thicklines \path(220,209)(240,209)
\thicklines \path(1436,209)(1416,209)
\thicklines \path(220,304)(240,304)
\thicklines \path(1436,304)(1416,304)
\put(198,304){\makebox(0,0)[r]{$1$}}
\thicklines \path(220,400)(240,400)
\thicklines \path(1436,400)(1416,400)
\thicklines \path(220,495)(240,495)
\thicklines \path(1436,495)(1416,495)
\put(198,495){\makebox(0,0)[r]{$2$}}
\thicklines \path(220,591)(240,591)
\thicklines \path(1436,591)(1416,591)
\thicklines \path(220,686)(240,686)
\thicklines \path(1436,686)(1416,686)
\put(198,686){\makebox(0,0)[r]{$3$}}
\thicklines \path(220,782)(240,782)
\thicklines \path(1436,782)(1416,782)
\thicklines \path(220,877)(240,877)
\thicklines \path(1436,877)(1416,877)
\put(198,877){\makebox(0,0)[r]{$4$}}
\thicklines \path(220,113)(220,133)
\thicklines \path(220,877)(220,857)
\put(220,68){\makebox(0,0){$0$}}
\thicklines \path(463,113)(463,133)
\thicklines \path(463,877)(463,857)
\put(463,68){\makebox(0,0){$0.2$}}
\thicklines \path(706,113)(706,133)
\thicklines \path(706,877)(706,857)
\put(706,68){\makebox(0,0){$0.4$}}
\thicklines \path(950,113)(950,133)
\thicklines \path(950,877)(950,857)
\put(950,68){\makebox(0,0){$0.6$}}
\thicklines \path(1193,113)(1193,133)
\thicklines \path(1193,877)(1193,857)
\put(1193,68){\makebox(0,0){$0.8$}}
\thicklines \path(1436,113)(1436,133)
\thicklines \path(1436,877)(1436,857)
\put(1436,68){\makebox(0,0){$1.0$}}
\thicklines \path(220,113)(1436,113)(1436,877)(220,877)(220,113)
\put(-145,495){\makebox(0,0)[l]{\shortstack{$\left| \frac{V_{ud}^\ast \, V_{ub}}{V_{td}^\ast \, V_{tb}}	\, {C_9}^{\rm CP}\right|$}}}
\put(828,23){\makebox(0,0){$\hat{s}$}}
\thicklines \path(220,638)(220,638)(220,638)(220,638)(220,637)(220,637)(220,637)(220,637)(220,637)(220,637)(220,637)(220,637)(220,637)(220,637)(220,637)(220,637)(220,637)(220,637)(220,637)(220,637)(220,637)(220,637)(220,637)(220,637)(220,637)(220,637)(220
,637)(220,637)(220,637)(220,637)(220,637)(220,637)(220,638)(220,640)(220,641)(220,642)(220,644)(220,645)(220,647)(220,649)(220,650)(220,652)(220,655)(220,657)(220,659)(220,662)(220,665)(220,669)(220,673)(220,677)(220,683)
\thicklines \path(220,683)(220,690)(220,694)(220,700)(220,703)(220,707)(220,710)(220,713)(220,715)(220,718)(220,719)(220,721)(220,723)(220,729)
\thicklines \path(220,729)(220,729)(220,583)(220,557)(220,543)(220,533)(220,525)(220,519)(220,509)(220,502)(220,496)(221,486)(221,479)(221,474)(221,465)(222,459)(222,454)(223,447)(223,445)(223,443)(224,441)(224,440)(224,439)(225,438)(225,437)(225,437)(225,437)(225,437)(225,437)(226,437)(226,437)(226,437)(226,436)(226,436)(226,436)(226,436)(226,436)(226,436)(226,436)(226,436)(226,436)(226,436)(226,437)(226,437)(226,437)(226,437)(226,437)(227,437)(227,438)(228,439)(228,441)
\thicklines \path(228,441)(229,443)(230,449)(232,456)(233,466)(234,478)(236,492)(237,509)(238,530)(239,554)(241,584)(242,619)(243,661)(245,711)(246,768)(247,831)(248,877)
\thicklines \path(253,877)(254,820)(254,766)(256,635)(258,540)(260,469)(262,415)(264,375)(266,344)(268,320)(270,302)(272,287)(274,276)(276,267)(280,254)(282,249)(284,246)(286,243)(288,240)(290,238)(292,236)(296,233)(300,231)(304,230)(308,229)(312,228)(316,228)(320,227)(324,227)(328,227)(332,226)(340,226)(348,226)(356,225)(360,225)(364,225)(368,225)(371,225)(373,225)(375,225)(377,225)(379,225)(380,225)(381,225)(382,225)(383,225)(384,225)(384,225)(385,225)(385,225)(386,225)
\thicklines \path(386,225)(386,225)(387,225)(387,225)(388,225)(388,225)(389,225)(389,225)(390,225)(391,225)(393,225)(395,225)(399,225)(403,225)(407,225)(411,225)(419,226)(427,226)(435,226)(443,226)(459,227)(475,229)(491,230)(507,233)(523,236)(538,240)(554,245)(570,252)(586,262)(594,268)(602,276)(610,285)(618,296)(626,311)(628,315)(630,320)(631,323)(632,327)(632,328)(633,331)(633,333)(634,339)
\thicklines \path(634,339)(634,339)(634,338)(634,337)(634,337)(634,337)(635,336)(635,336)(635,336)(635,336)(635,336)(635,336)(635,336)(635,336)(635,336)(636,336)(636,336)(636,336)(636,336)(636,336)(636,336)(636,336)(637,336)(637,336)(637,336)(638,336)(639,337)(640,337)(641,339)(645,344)(649,350)(653,357)(657,366)(661,376)(665,389)(668,403)(672,420)(676,441)(680,465)(684,495)(688,531)(692,577)(695,635)(699,711)(701,759)(703,815)(705,877)
\thicklines \path(739,877)(739,854)(742,735)(744,646)(746,577)(748,522)(753,439)(757,379)(761,335)(766,300)(770,272)(779,229)(788,199)(796,176)(801,166)(805,158)(810,151)(812,149)(814,146)(816,144)(817,143)(818,143)(820,142)(820,142)(821,142)(821,142)(821,142)(822,142)(822,142)(822,142)(823,141)(823,141)(823,141)(823,141)(824,141)(824,142)(824,142)(824,142)(825,142)(826,142)(826,142)(827,142)(829,144)(832,145)(836,150)(840,155)(849,169)(858,185)(867,205)(876,230)(884,263)
\thicklines \path(884,263)(889,283)(893,308)(898,337)(902,374)(906,422)(911,486)(913,527)(915,576)(917,636)(919,712)(922,809)(923,877)
\thicklines \path(944,877)(944,831)(945,700)(947,604)(948,531)(949,473)(951,426)(952,385)(954,350)(955,318)(956,289)(958,263)(959,238)(960,227)(960,217)(961,209)(961,206)(962,205)(962,204)(962,203)(962,203)(962,203)(962,203)(962,203)(962,203)(962,203)(962,203)(962,203)(962,203)(962,203)(963,203)(963,203)(963,203)(963,204)(963,204)(963,205)(963,207)(964,211)(964,217)(965,232)(965,253)(966,279)(967,344)(969,417)(970,492)
\thicklines \path(970,303)(970,303)(971,313)(971,317)(971,320)(971,321)(972,322)(972,323)(972,323)(972,324)(972,324)(972,324)(972,324)(973,324)(973,324)(973,324)(973,323)(973,322)(974,321)(975,313)(979,274)(984,245)(989,223)(993,206)(998,192)(1003,179)(1007,168)(1012,158)(1016,148)(1021,137)(1026,127)(1028,122)(1029,119)(1030,116)(1031,115)(1031,115)(1031,114)(1031,114)(1032,114)(1032,114)(1032,114)(1032,115)(1033,116)(1035,122)(1040,135)(1044,151)(1049,170)(1054,193)(1058,223)(1063,264)(1068,321)
\thicklines \path(1068,321)(1072,400)(1077,499)(1081,612)
\thicklines \path(1081,289)(1081,289)(1096,250)(1104,225)(1107,216)(1111,208)(1118,196)(1126,187)(1133,180)(1141,175)(1155,166)(1170,160)(1185,155)(1200,152)(1214,148)(1229,146)(1244,144)(1259,142)(1273,140)(1288,138)(1303,137)(1318,136)(1333,135)(1347,134)(1362,133)(1377,132)(1392,131)(1406,131)(1421,130)(1436,130)
\thinlines \path(220,544)(220,544)(220,544)(220,543)(220,543)(220,543)(220,543)(220,543)(220,543)(220,542)(220,542)(220,542)(220,542)(220,542)(220,542)(220,542)(220,542)(220,542)(220,542)(220,542)(220,542)(220,542)(220,542)(220,542)(220,542)(220,542)(220,542)(220,542)(220,542)(220,542)(220,543)(220,544)(220,545)(220,546)(220,547)(220,549)(220,550)(220,552)(220,554)(220,556)(220,557)(220,560)(220,562)(220,564)(220,567)(220,570)(220,574)(220,578)(220,582)(220,588)(220,594)
\thinlines \path(220,594)(220,599)(220,604)(220,608)(220,612)(220,615)(220,618)(220,620)(220,622)(220,624)(220,626)(220,628)(220,634)
\thinlines \path(220,634)(220,634)(221,394)(221,370)(222,357)(222,348)(223,335)(224,330)(224,326)(226,313)(228,308)(229,304)(233,292)(235,288)(237,284)(242,278)(246,273)(254,265)(263,259)(272,255)(289,248)(306,243)(323,239)(341,236)(358,234)(375,232)(392,230)(410,229)(427,227)(444,227)(453,226)(461,226)(470,226)(474,226)(479,226)(483,225)(487,225)(492,225)(494,225)(495,225)(496,225)(497,225)(498,225)(499,225)(499,225)(500,225)(500,225)(501,225)(501,225)(502,225)(502,225)
\thinlines \path(502,225)(503,225)(503,225)(504,225)(504,225)(506,225)(507,225)(509,225)(511,225)(513,225)(517,225)(522,226)(530,226)(539,226)(548,226)(556,227)(565,227)(573,228)(582,229)(591,230)(599,231)(608,233)(612,234)(617,235)(621,237)(623,238)(625,239)(627,240)(629,242)(631,243)(632,244)(632,245)(633,246)(633,247)(634,250)
\thinlines \path(634,250)(634,250)(635,243)(636,240)(638,236)(642,230)(646,225)(650,222)(667,210)(684,202)(701,195)(734,185)(767,177)(801,170)(834,165)(868,161)(901,157)(935,154)(968,151)(1001,148)(1035,146)(1068,144)(1102,142)(1135,140)(1169,139)(1202,137)(1235,136)(1269,135)(1302,134)(1336,133)(1369,132)(1403,131)(1436,130)
\thinlines \dashline[-10]{25}(221,697)(221,648)(222,621)(222,601)(223,575)(224,564)(224,556)(226,530)(228,520)(229,512)(233,487)(235,478)(237,470)(242,457)(246,447)(254,431)(263,419)(272,410)(289,396)(306,386)(323,378)(341,372)(358,368)(375,364)(392,361)(410,358)(427,356)(436,356)(444,355)(453,354)(461,354)(466,354)(470,354)(474,354)(476,354)(479,353)(481,353)(482,353)(483,353)(484,353)(485,353)(485,353)(486,353)(486,353)(487,353)(487,353)(488,353)(488,353)(489,353)(489,353)
\thinlines \dashline[-10]{25}(490,353)(490,353)(492,353)(493,353)(494,353)(496,353)(498,354)(500,354)(504,354)(509,354)(513,354)(522,354)(530,355)(539,356)(548,357)(556,358)(565,359)(573,361)(582,363)(591,365)(599,368)(608,372)(612,374)(617,377)(621,381)(623,383)(625,385)(627,388)(629,391)(631,393)(632,396)(632,398)(633,400)(633,403)(634,409)
\thinlines \dashline[-10]{25}(634,409)(635,395)(636,389)(638,381)(642,369)(646,360)(650,352)(667,329)(684,312)(701,298)(734,277)(767,261)(801,248)(834,237)(868,228)(901,220)(935,213)(968,207)(1001,202)(1035,197)(1068,193)(1102,189)(1135,186)(1169,183)(1202,180)(1235,177)(1269,175)(1302,172)(1336,170)(1369,168)(1403,167)(1436,165)
\thinlines \dashline[-10]{25}(220,344)(220,343)(220,342)(220,342)(220,341)(220,341)(220,341)(220,340)(220,340)(220,340)(220,339)(220,339)(220,339)(220,339)(220,338)(220,338)(220,338)(220,338)(220,338)(220,338)(220,338)(220,338)(220,338)(220,338)(220,338)(220,338)(220,338)(220,338)(220,338)(220,338)(220,338)(220,338)(220,338)(220,338)(220,338)(220,338)(220,338)(220,338)(220,338)(220,338)(220,338)(220,338)(220,338)(220,338)(220,338)(220,338)(220,338)(220,339)(220,339)(220,340)
\thinlines \dashline[-10]{25}(220,340)(220,341)(220,342)(220,343)(220,344)(220,345)(220,346)(220,347)(220,348)(220,350)(220,352)(220,354)(220,356)(220,360)(220,362)(220,364)(220,366)(220,368)(220,369)(220,371)(220,372)(220,373)(220,373)(220,374)(220,375)(220,378)
\thinlines \dashline[-10]{25}(220,378)(221,256)(221,244)(222,237)(222,233)(223,226)(224,224)(224,221)(226,215)(228,213)(229,211)(233,204)(235,202)(237,200)(242,197)(246,194)(254,190)(263,187)(272,185)(289,182)(306,179)(323,177)(341,175)(358,174)(375,173)(392,172)(410,171)(427,170)(444,170)(461,170)(479,169)(487,169)(496,169)(500,169)(504,169)(509,169)(513,169)(517,169)(520,169)(521,169)(522,169)(523,169)(523,169)(524,169)(524,169)(525,169)(525,169)(526,169)(527,169)(527,169)
\thinlines \dashline[-10]{25}(528,169)(528,169)(529,169)(530,169)(531,169)(532,169)(535,169)(537,169)(539,169)(543,169)(548,169)(556,169)(561,169)(565,169)(573,169)(582,170)(591,170)(599,170)(608,171)(612,171)(617,172)(621,173)(623,173)(625,173)(627,174)(629,174)(631,175)(632,175)(632,176)(633,176)(633,177)(634,178)
\thinlines \dashline[-10]{25}(634,178)(635,174)(636,173)(638,171)(642,168)(646,166)(650,164)(667,158)(684,154)(701,151)(734,146)(767,142)(801,139)(834,136)(868,134)(901,132)(935,131)(968,130)(1001,128)(1035,127)(1068,127)(1102,126)(1135,125)(1169,124)(1202,124)(1235,123)(1269,123)(1302,122)(1336,122)(1369,122)(1403,121)(1436,121)
\thinlines \dashline[-10]{25}(220,332)(220,330)(220,330)(220,329)(220,328)(220,328)(220,328)(220,327)(220,327)(220,327)(220,326)(220,326)(220,326)(220,326)(220,326)(220,325)(220,325)(220,325)(220,325)(220,325)(220,325)(220,325)(220,325)(220,325)(220,325)(220,325)(220,325)(220,325)(220,325)(220,325)(220,325)(220,325)(220,325)(220,325)(220,325)(220,325)(220,325)(220,325)(220,325)(220,325)(220,325)(220,325)(220,325)(220,325)(220,325)(220,325)(220,325)(220,325)(220,325)(220,326)
\thinlines \dashline[-10]{25}(220,326)(220,327)(220,327)(220,328)(220,329)(220,330)(220,331)(220,332)(220,333)(220,334)(220,335)(220,337)(220,338)(220,341)(220,343)(220,346)(220,348)(220,351)(220,353)(220,355)(220,356)(220,357)(220,358)(220,359)(220,360)(
220,361)(220,362)(220,365)
\thinlines \dashline[-10]{25}(220,365)(221,257)(221,245)(222,239)(222,234)(223,227)(224,225)(224,223)(226,216)(228,214)(229,212)(233,206)(235,204)(237,202)(242,199)(246,196)(254,192)(263,190)(272,187)(289,184)(306,182)(323,180)(341,179)(358,178)(375,177)(392,176)(410,176)(418,175)(427,175)(436,175)(444,175)(453,175)(457,175)(459,175)(461,175)(464,175)(465,175)(466,175)(467,175)(468,175)(468,175)(469,175)(469,175)(470,175)(471,175)(471,175)(472,175)(472,175)(473,175)(473,175)
\thinlines \dashline[-10]{25}(474,175)(474,175)(475,175)(476,175)(479,175)(481,175)(483,175)(487,175)(492,175)(496,175)(504,175)(513,175)(522,175)(530,176)(548,176)(565,177)(573,177)(582,178)(591,179)(599,179)(608,180)(612,181)(617,182)(621,182)(623,183)(625,184)(627,184)(629,185)(631,186)(632,186)(632,187)(633,187)(633,188)(634,190)
\thinlines \dashline[-10]{25}(634,190)(635,186)(636,185)(638,183)(642,180)(646,178)(650,176)(667,171)(684,167)(701,164)(734,159)(767,155)(801,152)(834,150)(868,147)(901,146)(935,144)(968,143)(1001,141)(1035,140)(1068,139)(1102,138)(1135,138)(1169,137)(1202,136)(1235,136)(1269,135)(1302,134)(1336,134)(1369,134)(1403,133)(1436,133)
\end{picture}
        \end{center}
        \caption{The magnitude of ${C_9}^{\rm CP}$ 
	with (solid thick curve) and 
	without (solid thin curve) resonances for 
	$(\gamma,x) = ({48.6}^o,0.778)$. The 
	upper, middle and lower dashed curves show 
	the magnitude without resonances for same $x$ and 
	$\gamma = 0^o, {-90}^o, {90}^o$.}
        \label{fig:c9cp}
\end{figure}
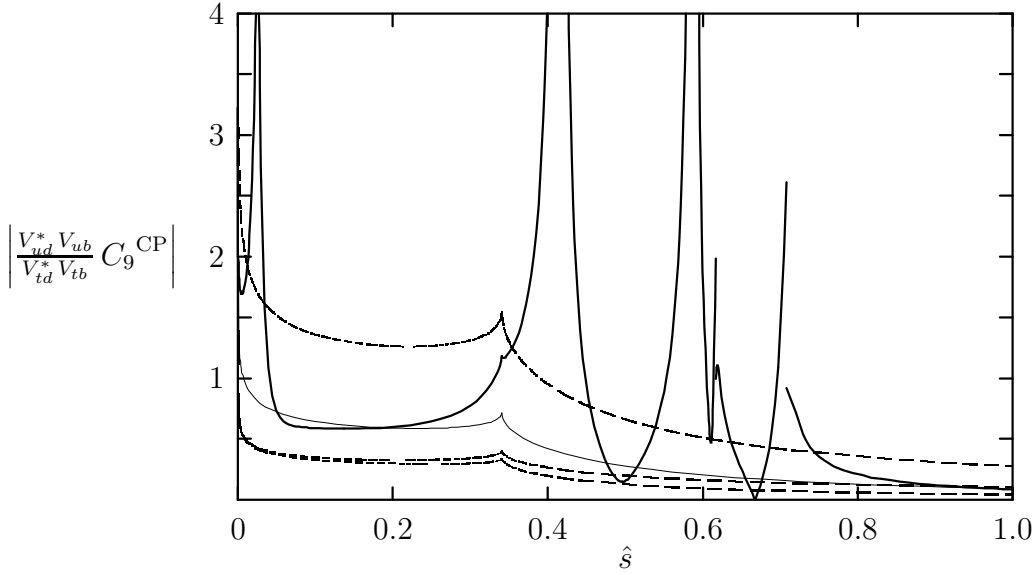

Now let me derive some relations in the CKM unitarity triangle and 
give numerical calculation for the auxiliary functions 
defined above. Especially it is worthwhile to see how 
large the contribution of ${C_9}^{\rm CP}$.
Using Wolfenstein parametrization \cite{wolfenstein}, one can 
rewrite the CKM factor as
\begin{equation}
	\frac{V_{uq}^\ast \, V_{ub}}{V_{tq}^\ast \, V_{tb}} \sim 
	\left\{
	\begin{array}{lcl}
		\frac{\displaystyle 
		r \left( e^{-i \gamma} - r \right)}{
		\displaystyle 1 + r^2 - 2 \, r \, \cos \gamma}	& , & q = d \\
		\lambda^2 \, r \, e^{-i \gamma} & , & q = s 
	\end{array}
	\right. \; . 
	\label{eq:ckmfactor}
\end{equation}
As mentioned before, from Eqs. (\ref{eq:ratiockms}) and 
(\ref{eq:ckmfactor}) it is obvious that in the 
$b \rightarrow s \, \l^+ \, \l^-$ decay, $u\bar{u}$ loop and
then ${C_9}^{\rm CP}$ is less important and negligible. 
In general one must treat $r$ and $\gamma$ as free parameters, 
while $x$ must be determined by the data of $B_d^0-\bar{B}_d^0$ mixing. 
However, in the SM the unitarity triangle is satisfied in 
a good aproximation, so one can relate $r$ and $x$ each other 
as below
\begin{equation}
	r = \sqrt{| x^2 - \sin^2 \gamma |} + \cos \gamma \; .
	\label{eq:r}
\end{equation}
Here $x$ is determined by the experimental value of $x_d$ in the 
$B_d^0-\bar{B}_d^0$ mixing, that is
\begin{equation}
	x = \left[ \frac{x_d}{
	{{G_F}^2}/{(6 \, \pi^2)} \, m_{B_d} \, {M_W}^2 \, 
		\tau_{B_d} \, \eta_{\rm QCD} \, {f_{B_d}}^2 \, B_{B_d} \,
		\left| F_{\Delta B=2} \right|^2} \right]^{1/2} \; ,
	\label{eq:x}
\end{equation}
with \cite{bb}
\begin{equation}
	F_{\Delta B=2} = 
	\frac{4 - 15 \, x_t + 12 \, {x_t}^2 - {x_t}^3 
		- 6 \, {x_t}^2 \, \ln x_t}{
		4 \, (1 - x_t)^3} \; ,
	\label{eq:fd2}
\end{equation}
and $x_t = \left( {m_t}/{m_W} \right)^2$. 
For $m_{B_d}$ and $\tau_{B_d}$, I use $m_{B^0}$ and $\tau_{B^0}$.

The magnitude of $\left( {V_{ud}^\ast \, V_{ub}}/{V_{td}^\ast \, V_{tb}} 
\right) {C_9}^{\rm CP}$ as a function of $\s$ is depicted 
in Fig. \ref{fig:c9cp}. Here I use the following values for 
the Wilson coefficients
\[
	C_1 = -0.2404  , \; 
	C_2 = 1.1032  , \;
	C_3 = 0.0107  , \;
	C_4 = -0.0249  , \; 
	C_5 = 0.0072  , 
\]
\[
	C_6 = -0.03024 , \;
	{C_7}^{\rm eff} = -0.3109 , \;
	C_8 = -0.1478 , 
	{C_9}^{\rm NLO} = 4.1990 , \; 
	C_{10} = -4.5399 , 
\]
that are obtained by using the central values in 
Tab. (\ref{tab:parameters}), The solid curves show the magnitude for 
$(\gamma,x) = ({48.6}^o,0.778)$ that is the best fit in the SM  up to now 
and equivalent to $(\rho,\eta) = (0.3,0.34)$. The upper, middle and 
lower dashed curves show the magnitude without LD effects 
with $\gamma = 0^o, {-90}^o, {90}^o$. 
It is obvious that the contribution is significant, about 
$\sim 20\%$ of ${C_9}^{\rm NLO}$ at low $\s$ region 
($\s < 0.4$). 

Now, I am ready to analyse the decay rate and asymmetries in the 
channel.

\section{Decay rate and asymmetries}
\label{sec:dra}

The double differential decay rate for semi-leptonic 
$B \rightarrow X_q \,\l^+ \, \l^-$ decay, involving the lepton and 
light quark masses, is expressed as
\begin{eqnarray}
	\frac{{\rm d}^2{\cal B}(\s,z)}{{\rm d}\s {\rm d}z} & = & 
	{\cal B}_o \sqrt{1 - \frac{4 \, \ml^2}{\s}} \, \u(\s)\, \left\{
	4 \, \left[ \left| {C_9}^{\rm eff} \right|^2 
		- \left| C_{10} \right|^2 \right] 
		\ml^2 \, \left[ 1 - \s + \mq^2 \right]
	\right.
	\nonumber \\
	& & \left.
	+ \left[ \left| {C_9}^{\rm eff} \right|^2 
		+ \left| C_{10} \right|^2 \right]
		\left[ (1 - \mq^2)^2 - \s^2 - {\u(\s)}^2 \, 
		\left( 1 - \frac{6 \, \ml^2}{\s} \right) \, z^2 \right]
	\right.
	\nonumber \\
	& & \left.
	+ 4 \, \left| {C_7}^{\rm eff} \right|^2	\, 
		\frac{1 + {2 \, \ml^2}/\s}{\s}
	\right.
	\nonumber \\
	& & \left. \; \; \; \;
		\times \left[ 1 - \mq^2 - \mq^4 + \mq^6 
		- \s \, (8 \, \mq^2 + \s + \mq^2 \, \s) 
		+ {\u(\s)}^2 \, (1 + \mq^2) \, z^2 \right] 
	\right.
	\nonumber \\
	& & \left.
	- 8 \, {\rm Re} \left( {C_9}^{\rm eff} \right)^\ast \, {C_7}^{\rm eff}\,
		\left[ 1 + \frac{2 \, \ml^2}{\s} \right] \, 
		\left[ \s \, (1 + \mq^2) - (1 - \mq^2)^2 \right]
	\right.
	\nonumber \\
	& & \left.
	+ 4 \, C_{10} \, \left[ {\rm Re} \left( {C_9}^{\rm eff} \right)^\ast \, \s
		+ 2 \, {C_7}^{\rm eff} \, (1 + \mq^2) \right] \, \u(\s) \, z 
	\right\} \; , 
	\label{eq:dddr}
\end{eqnarray}
where $\u(\s) = \sqrt{[\s - (1 + \mq)^2][\s - (1 - \mq)^2]}$,
$z = \cos \theta$ is the angle of $\l^+$ measured with 
respect to the $b-$quark direction in the dilepton CM system
and the normalization factor, 
\begin{equation}
	{\cal B}_o = {\cal B} (B \rightarrow X_c \, \l \, \bar{\nu}) 
	\frac{3 \ \alpha^2}{16 \, \pi^2} \, 
	\frac{\left| V_{tq}^\ast \, V_{tb} \right|^2}{
		\left| V_{cb} \right|^2} \, 
	\frac{1}{f(\mc) \, \kappa(\mc)} \; ,
	\label{eq:bo}
\end{equation}
is to reduce the uncertainty due to $b-$quark mass. In the 
preceding notation, the CKM factor would read 
${\left| V_{td}^\ast \, V_{tb} \right|^2}/{\left| V_{cb} \right|^2} 
\sim \lambda^2 \, x^2$.
$f(\mc)$ is the phase space function for 
$\Gamma(B \rightarrow X_c \, \l \, \nu)$ in parton model, 
while $\kappa(\mc)$ accounts the $O(\alpha_s)$ QCD correction 
to the decay. Writing both functions explicitly, 
\begin{eqnarray}
	f(\mc) & = & 1 - 8 \, \mc^2 + 8 \, \mc^6 - \mc^8 
		- 24 \, \mc^4 \, \ln \mc \; , \\
	\label{eq:fmc}
	\kappa(\mc) & = & 1 - 
	\frac{2 \, \alpha_s(m_b)}{3 \, \pi} \left[
	\frac{3}{2} + \left( \pi^2 - \frac{31}{4} \right) 
	(1 - \mc)^2 \right] \; ,
	\label{eq:kappa}
\end{eqnarray}
and using the values in Tab. \ref{tab:parameters}, 
$f(\mc) = 0.542$ and $\kappa(\mc) = 0.885$.

\subsection{\bf Decay rate}
\label{sec:dr}

\begin{figure}[t]
        \begin{minipage}[t]{\minitwocolumn}
        \begin{center}
\setlength{\unitlength}{0.240900pt}
\begin{picture}(1049,900)(0,0)
\thicklines \path(220,189)(240,189)
\thicklines \path(985,189)(965,189)
\thicklines \path(220,266)(240,266)
\thicklines \path(985,266)(965,266)
\put(198,266){\makebox(0,0)[r]{$1$}}
\thicklines \path(220,342)(240,342)
\thicklines \path(985,342)(965,342)
\thicklines \path(220,419)(240,419)
\thicklines \path(985,419)(965,419)
\put(198,419){\makebox(0,0)[r]{$2$}}
\thicklines \path(220,495)(240,495)
\thicklines \path(985,495)(965,495)
\thicklines \path(220,571)(240,571)
\thicklines \path(985,571)(965,571)
\put(198,571){\makebox(0,0)[r]{$3$}}
\thicklines \path(220,648)(240,648)
\thicklines \path(985,648)(965,648)
\thicklines \path(220,724)(240,724)
\thicklines \path(985,724)(965,724)
\put(198,724){\makebox(0,0)[r]{$4$}}
\thicklines \path(220,801)(240,801)
\thicklines \path(985,801)(965,801)
\thicklines \path(220,877)(240,877)
\thicklines \path(985,877)(965,877)
\put(198,877){\makebox(0,0)[r]{$5$}}
\thicklines \path(220,113)(220,133)
\thicklines \path(220,877)(220,857)
\put(220,68){\makebox(0,0){$0$}}
\thicklines \path(373,113)(373,133)
\thicklines \path(373,877)(373,857)
\put(373,68){\makebox(0,0){$0.2$}}
\thicklines \path(526,113)(526,133)
\thicklines \path(526,877)(526,857)
\put(526,68){\makebox(0,0){$0.4$}}
\thicklines \path(679,113)(679,133)
\thicklines \path(679,877)(679,857)
\put(679,68){\makebox(0,0){$0.6$}}
\thicklines \path(832,113)(832,133)
\thicklines \path(832,877)(832,857)
\put(832,68){\makebox(0,0){$0.8$}}
\thicklines \path(985,113)(985,133)
\thicklines \path(985,877)(985,857)
\put(985,68){\makebox(0,0){$1.0$}}
\thicklines \path(220,113)(985,113)(985,877)(220,877)(220,113)
\put(0,495){\makebox(0,0)[l]{\shortstack{$\frac{{\rm d}{\cal B}}{{\rm d}\hat{s}}$}}}
\put(220,920){\makebox(0,0)[l]{\shortstack{$(\times {10}^{-7})$}}}
\put(602,23){\makebox(0,0){$\hat{s}$}}
\thicklines \path(241,877)(241,834)(241,803)(242,786)(242,775)(242,765)(243,751)(243,739)(244,730)(245,716)(248,694)(250,676)(255,645)(260,621)(270,585)(275,572)(280,561)(290,543)(300,530)(310,519)(320,510)(330,503)(340,496)(350,490)(360,485)(370,480)(380
,475)(390,470)(400,465)(410,461)(420,457)(430,453)(440,449)(445,448)(450,447)(455,445)(458,445)(459,445)(460,445)(462,445)(462,444)(463,444)(463,444)(464,444)(464,444)(464,444)(465,444)(465,444)(465,444)(466,444)(466,444)
\thicklines \path(466,444)(466,444)(467,444)(467,444)(467,444)(468,444)(468,444)(469,444)(470,445)(472,445)(473,445)(474,446)(475,446)(477,447)(477,447)(478,448)(478,448)(479,449)(479,449)(480,450)(480,451)(480,452)
\thicklines \path(480,452)(480,452)(480,453)(480,453)(481,454)(481,454)(481,455)(482,455)(483,457)(485,459)(488,461)(490,462)(492,464)(495,467)(497,469)(500,473)(502,477)(505,482)(507,488)(509,496)(512,507)(514,521)(517,540)(518,552)(519,567)(520,584)(521
,606)(523,632)(524,665)(525,708)(526,763)(528,839)(528,877)
\thicklines \path(543,877)(543,874)(543,787)(544,718)(544,663)(545,582)(545,551)(545,526)(546,486)(547,457)(547,435)(548,419)(549,406)(549,396)(550,389)(551,382)(552,377)(552,373)(553,370)(554,367)(555,363)(556,360)(558,357)(559,356)(561,355)(563,353)(566
,352)(572,350)(577,348)(583,346)(588,344)(594,341)(599,339)(605,336)(610,334)(616,331)(621,329)(624,328)(627,328)(628,327)(630,327)(630,327)(631,327)(632,327)(632,327)(632,327)(633,327)(633,327)(633,327)(633,327)(633,327)
\thicklines \path(633,327)(633,327)(634,327)(634,327)(634,327)(634,327)(634,327)(634,327)(635,327)(635,327)(636,327)(636,327)(637,327)(637,327)(638,327)(639,328)(641,328)(642,329)(643,330)(645,332)(646,333)(648,335)(649,338)(650,342)(652,346)(653,351)(655
,358)(656,367)(657,379)(659,396)(659,406)(660,418)(661,451)(662,474)(663,502)(664,538)(664,585)(665,649)(665,690)(666,738)(666,867)(666,877)
\thicklines \path(673,877)(673,816)(673,684)(673,634)(673,591)(673,524)(674,474)(674,436)(674,406)(674,383)(675,364)(675,349)(675,337)(675,318)(676,305)(676,295)(677,288)(677,283)(678,279)(678,275)(678,273)(679,271)(679,270)(680,269)(680,268)(680,268)(681
,267)(681,267)(681,267)(681,267)(681,267)(681,267)(682,267)(682,267)(682,267)(682,267)(682,267)(682,267)(682,267)(682,267)(682,267)(682,267)(682,267)(682,267)(682,267)(682,267)(682,267)(682,267)(682,267)(682,267)(682,267)
\thicklines \path(682,267)(682,267)(683,267)(683,267)(683,267)(684,268)(684,268)(685,269)(685,270)(685,271)(686,273)(686,275)(687,277)(687,279)(688,283)(688,291)(689,302)(690,317)(691,331)(692,342)
\thicklines \path(692,283)(692,283)(693,275)(694,270)(695,265)(696,262)(696,259)(697,257)(698,256)(699,253)(701,252)(704,249)(706,247)(709,245)(712,244)(715,242)(718,240)(721,239)(724,237)(727,236)(730,234)(733,233)(734,233)(736,232)(736,232)(737,232)(738
,232)(738,232)(739,232)(739,232)(739,232)(739,232)(740,232)(740,232)(740,232)(740,232)(740,232)(740,232)(740,232)(740,232)(740,232)(740,232)(741,232)(741,232)(741,232)(741,232)(741,232)(741,232)(742,232)(742,232)(743,232)
\thicklines \path(743,232)(744,232)(744,232)(745,233)(746,233)(747,234)(747,234)(749,236)(750,239)(752,242)(753,246)(755,252)(756,259)(759,276)(762,290)
\thicklines \path(762,230)(762,230)(767,215)(769,209)(771,204)(774,201)(776,198)(781,194)(790,187)(799,180)(808,174)(818,169)(827,163)(836,158)(846,153)(855,148)(864,143)(873,139)(883,135)(892,131)(901,127)(911,124)(920,121)(929,118)(934,116)(936,116)(939
,115)(940,114)(940,114)(941,114)(941,114)(941,113)
\thinlines \path(239,877)(242,823)(244,778)(247,743)(253,689)(258,652)(263,623)(269,602)(274,584)(285,559)(296,540)(307,526)(318,515)(328,506)(339,498)(350,491)(361,484)(372,477)(383,471)(394,465)(404,459)(415,453)(426,448)(437,442)(448,436)(459,430)(464,
427)(469,425)(472,424)(474,423)(475,423)(476,423)(476,423)(477,422)(477,422)(478,422)(478,422)(478,422)(479,422)(479,422)(479,422)(480,423)(480,423)(480,424)
\thinlines \path(480,424)(480,424)(481,426)(482,426)(482,426)(483,427)(484,427)(484,427)(485,427)(486,426)(487,426)(488,426)(491,425)(501,420)(522,406)(543,391)(564,375)(585,357)(606,340)(628,322)(649,303)(670,285)(691,267)(712,250)(733,233)(754,216)(775,
200)(796,185)(817,171)(838,158)(859,147)(880,136)(901,128)(922,120)(932,117)(938,115)(939,115)(940,114)(940,114)(941,114)
\thinlines \dashline[-10]{25}(247,868)(253,809)(258,767)(263,735)(269,710)(274,689)(285,658)(296,634)(307,615)(318,600)(328,586)(339,574)(350,563)(361,552)(372,542)(383,532)(394,522)(404,512)(415,501)(426,491)(437,481)(448,470)(459,459)(464,453)(469,446)(
472,443)(475,439)(476,437)(478,435)(479,432)(479,432)(480,431)(480,430)(480,427)
\thinlines \dashline[-10]{25}(480,427)(501,408)(522,392)(543,377)(564,361)(585,344)(606,328)(628,311)(649,294)(670,277)(691,260)(712,243)(733,227)(754,212)(775,197)(796,182)(817,169)(838,157)(859,145)(880,136)(901,127)(922,120)(932,117)(938,115)(939,115)(
940,114)(940,114)(941,114)
\thinlines \dashline[-10]{25}(242,838)(244,794)(247,759)(253,706)(258,669)(263,641)(269,620)(274,603)(285,578)(296,560)(307,547)(318,536)(328,527)(339,519)(350,512)(361,505)(372,499)(383,493)(394,487)(404,481)(415,475)(426,469)(437,463)(448,457)(459,451)(
464,449)(469,446)(472,445)(474,444)(475,444)(476,444)(476,443)(477,443)(477,443)(478,443)(478,443)(478,443)(479,443)(479,443)(479,443)(480,443)(480,444)(480,445)
\thinlines \dashline[-10]{25}(480,445)(501,433)(522,417)(543,399)(564,381)(585,363)(606,344)(628,325)(649,307)(670,288)(691,270)(712,252)(733,234)(754,217)(775,201)(796,186)(817,172)(838,159)(859,147)(880,137)(901,128)(922,120)(932,117)(938,115)(939,115)(
940,114)(940,114)(941,114)
\thinlines \dashline[-10]{25}(242,871)(244,827)(247,792)(253,739)(258,703)(263,675)(269,654)(274,638)(285,613)(296,596)(307,582)(318,572)(328,563)(339,556)(350,549)(361,542)(372,536)(383,530)(394,524)(404,517)(415,511)(426,505)(437,499)(448,493)(459,487)(
464,484)(469,481)(472,480)(474,480)(475,479)(476,479)(476,479)(477,479)(477,479)(477,479)(478,479)(478,479)(478,479)(479,479)(479,479)(479,479)(480,479)(480,480)(480,481)
\thinlines \dashline[-10]{25}(480,481)(501,455)(522,434)(543,413)(564,392)(585,372)(606,351)(628,331)(649,311)(670,292)(691,273)(712,254)(733,236)(754,219)(775,203)(796,187)(817,173)(838,159)(859,147)(880,137)(901,128)(922,120)(932,117)(938,115)(939,115)(
940,114)(940,114)(941,114)
\end{picture}
        \end{center}
        \end{minipage}
        \hfill
        \begin{minipage}[t]{\minitwocolumn}
        \begin{center}
\setlength{\unitlength}{0.240900pt}
\begin{picture}(1049,900)(0,0)
\thicklines \path(220,189)(240,189)
\thicklines \path(985,189)(965,189)
\thicklines \path(220,266)(240,266)
\thicklines \path(985,266)(965,266)
\thicklines \path(220,342)(240,342)
\thicklines \path(985,342)(965,342)
\thicklines \path(220,419)(240,419)
\thicklines \path(985,419)(965,419)
\thicklines \path(220,495)(240,495)
\thicklines \path(985,495)(965,495)
\put(198,495){\makebox(0,0)[r]{$1$}}
\thicklines \path(220,571)(240,571)
\thicklines \path(985,571)(965,571)
\thicklines \path(220,648)(240,648)
\thicklines \path(985,648)(965,648)
\thicklines \path(220,724)(240,724)
\thicklines \path(985,724)(965,724)
\thicklines \path(220,801)(240,801)
\thicklines \path(985,801)(965,801)
\thicklines \path(220,877)(240,877)
\thicklines \path(985,877)(965,877)
\put(198,877){\makebox(0,0)[r]{$2$}}
\thicklines \path(220,113)(220,133)
\thicklines \path(220,877)(220,857)
\put(220,68){\makebox(0,0){$0$}}
\thicklines \path(373,113)(373,133)
\thicklines \path(373,877)(373,857)
\put(373,68){\makebox(0,0){$0.2$}}
\thicklines \path(526,113)(526,133)
\thicklines \path(526,877)(526,857)
\put(526,68){\makebox(0,0){$0.4$}}
\thicklines \path(679,113)(679,133)
\thicklines \path(679,877)(679,857)
\put(679,68){\makebox(0,0){$0.6$}}
\thicklines \path(832,113)(832,133)
\thicklines \path(832,877)(832,857)
\put(832,68){\makebox(0,0){$0.8$}}
\thicklines \path(985,113)(985,133)
\thicklines \path(985,877)(985,857)
\put(985,68){\makebox(0,0){$1.0$}}
\thicklines \path(220,113)(985,113)(985,877)(220,877)(220,113)
\put(220,920){\makebox(0,0)[l]{\shortstack{$(\times {10}^{-6})$}}}
\put(602,23){\makebox(0,0){$\hat{s}$}}
\thicklines \path(241,877)(241,746)(241,707)(242,684)(242,653)(243,629)(243,611)(244,597)(245,587)(245,579)(247,567)(248,558)(250,545)(255,526)(260,511)(270,487)(280,470)(290,457)(300,447)(310,439)(320,431)(330,425)(340,419)(350,413)(360,408)(370,403)(380,398)(390,393)(400,387)(410,382)(420,377)(430,372)(440,366)(450,360)(460,355)(470,348)(480,342)
\thicklines \path(480,342)(480,342)(480,342)(480,343)(481,343)(481,343)(481,343)(481,343)(481,343)(481,343)(481,343)(481,343)(482,343)(482,343)(482,343)(482,343)(482,343)(482,343)(482,343)(482,343)(482,343)(482,343)(482,343)(482,343)(483,343)(483,343)(483,343)(483,343)(484,343)(485,343)(488,342)(490,341)(492,340)(495,339)(497,338)(500,337)(502,336)(505,335)(507,334)(509,333)(511,333)(512,333)(512,332)(513,332)(514,332)(514,332)(514,332)(514,332)(515,332)(515,332)(515,332)
\thicklines \path(515,332)(515,332)(515,332)(515,332)(515,332)(515,332)(515,332)(515,332)(515,332)(515,332)(516,332)(516,332)(516,332)(516,332)(516,332)(516,332)(517,332)(517,332)(517,332)(518,333)(518,333)(519,333)(520,334)(520,334)(521,335)(521,336)(522,337)(523,338)(523,339)(524,341)(525,343)(525,346)(526,349)(526,352)(527,357)(528,362)(528,369)(529,378)(529,388)(530,402)(531,419)(531,442)(532,472)(532,513)(533,539)(533,570)(533,607)(534,653)(534,709)(534,778)(535,865)
\thicklines \path(535,865)(535,877)
\thicklines \path(541,877)(542,800)(542,715)(542,651)(543,602)(543,563)(543,533)(544,507)(544,487)(545,455)(545,431)(546,413)(547,400)(547,389)(548,380)(549,366)(551,357)(552,349)(554,343)(555,339)(558,331)(561,326)(566,318)(572,311)(577,306)(583,301)(588,297)(594,293)(599,288)(605,284)(610,280)(616,276)(621,272)(627,268)(632,264)(638,260)(643,256)(649,252)(652,250)(653,250)(654,249)(655,249)(655,249)(656,249)(656,249)(656,249)(656,249)(656,249)(657,249)(657,249)(657,249)
\thicklines \path(657,249)(657,249)(657,249)(657,249)(658,249)(658,249)(658,249)(658,249)(659,249)(659,249)(659,249)(660,250)(660,250)(661,251)(661,252)(661,253)(662,255)(663,259)(663,261)(664,263)(664,266)(664,270)(665,274)(665,279)(665,285)(666,293)(666,303)(666,315)(667,331)(667,351)(667,378)(668,414)(668,437)(668,464)(668,497)(668,536)(669,584)(669,644)(669,719)(669,815)(669,877)
\thicklines \path(672,877)(672,788)(672,699)(672,630)(673,577)(673,534)(673,499)(673,471)(673,427)(673,394)(673,370)(674,352)(674,325)(674,316)(675,307)(675,295)(675,286)(676,278)(676,273)(677,268)(677,265)(678,259)(679,255)(680,251)(681,249)(682,247)(682,245)(683,243)(684,242)(685,242)(685,241)(685,241)(685,241)(686,241)(686,241)(686,241)(686,241)(686,241)(686,241)(686,241)(686,241)(686,241)(686,241)(686,241)(686,241)(686,241)(686,241)(686,241)(686,241)(686,241)(686,241)
\thicklines \path(686,241)(686,241)(686,241)(687,241)(687,241)(687,241)(687,241)(687,241)(687,241)(687,241)(688,241)(688,242)(688,243)(688,243)(689,245)(689,247)(690,249)(690,253)(691,260)(692,268)
\thicklines \path(692,246)(692,246)(695,239)(696,236)(698,233)(701,229)(704,226)(706,223)(709,221)(712,218)(715,216)(718,214)(721,212)(724,210)(727,208)(730,206)(733,205)(736,203)(739,201)(741,200)(743,199)(744,198)(746,198)(747,198)(747,197)(748,197)(748,197)(748,197)(748,197)(749,197)(749,197)(749,197)(749,197)(749,197)(749,197)(749,197)(749,197)(749,197)(749,197)(750,197)(750,197)(750,197)(750,197)(750,197)(750,197)(751,197)(751,198)(751,198)(752,198)(752,198)(753,199)
\thicklines \path(753,199)(754,199)(755,200)(756,203)(758,206)(759,210)(762,220)
\thicklines \path(762,197)(762,197)(771,187)(781,178)(790,172)(799,167)(808,162)(818,157)(827,153)(836,149)(846,144)(855,140)(864,137)(873,133)(883,130)(892,127)(901,124)(911,121)(920,119)(929,117)(934,116)(936,115)(939,114)(940,114)(940,114)(941,114)(941,114)(941,113)
\thicklines \dashline[-10]{25}(222,765)(223,681)(223,617)(224,568)(224,529)(225,470)(226,428)(227,396)(227,372)(228,353)(229,337)(230,325)(231,314)(231,306)(232,298)(233,292)(234,288)(235,284)(236,281)(236,278)(237,276)(238,274)(238,273)(238,272)(238,272)(239,272)(239,272)(239,272)(239,272)(239,272)(239,272)(239,272)(239,272)(239,272)(239,272)(239,272)(239,272)(239,272)(239,272)(239,272)(239,272)(239,272)(239,272)(239,273)(239,273)(239,274)(239,274)(239,276)(239,277)(240,279)
\thicklines \dashline[-10]{25}(240,280)(240,281)(240,282)(240,284)
\thicklines \dashline[-10]{25}(240,272)(240,273)(240,274)(240,276)(240,278)(240,279)(240,282)(240,284)(240,287)(240,290)(240,294)(240,298)(240,304)(240,309)(240,316)(240,324)(240,333)(240,343)(240,354)(240,366)(240,377)(240,388)(240,392)(240,396)(240,397)(240,399)(240,400)(240,400)(240,400)(240,401)(240,401)(240,401)(240,401)(240,401)(240,401)(240,401)(240,401)(240,401)(240,401)(240,401)(240,401)(240,401)(240,401)(240,401)(240,401)(240,401)(240,401)(240,401)(240,401)(240,401)
\thicklines \dashline[-10]{25}(240,401)(240,401)(240,401)(240,401)
\thicklines \dashline[-10]{25}(240,353)(241,259)(241,236)(241,229)(242,226)(242,221)(243,218)(243,215)(244,213)(245,211)(247,209)(248,207)(250,204)(255,200)(260,196)(270,190)(275,188)(280,187)(290,184)(300,182)(310,180)(320,178)(330,177)(340,176)(350,175)(360,174)(370,173)(380,172)(390,171)(400,170)(410,169)(420,168)(430,168)(440,167)(450,166)(460,165)(465,165)(468,165)(469,165)(470,165)(472,165)(472,165)(473,165)(473,165)(474,165)(474,165)(474,165)(475,165)(475,165)(475,165)
\thicklines \dashline[-10]{25}(476,165)(476,165)(476,165)(477,165)(477,165)(477,165)(477,165)(478,165)(478,165)(478,165)(479,165)(479,165)(479,165)(480,165)(480,165)(480,165)
\thicklines \dashline[-10]{25}(480,165)(480,165)(480,165)(481,166)(481,166)(481,166)(482,166)(482,166)(483,166)(485,166)(488,166)(490,167)(492,167)(495,167)(497,167)(500,167)(502,168)(505,168)(507,169)(509,170)(512,171)(514,172)(515,173)(517,175)(518,176)
(519,178)(520,180)(521,182)(523,185)(524,189)(525,195)(526,201)(528,211)(528,217)(529,224)(529,232)(530,243)(531,256)(531,273)(532,295)(532,324)(533,363)(533,388)(534,418)(534,454)(534,499)(535,554)(535,625)(535,716)(535,837)
\thicklines \dashline[-10]{25}(535,877)
\thicklines \dashline[-10]{25}(540,816)(540,696)(541,605)(541,534)(541,479)(541,434)(541,398)(542,343)(542,304)(542,276)(543,254)(543,238)(543,225)(544,215)(544,206)(545,194)(545,189)(545,185)(546,179)(547,174)(547,171)(548,168)(549,166)(549,164)(550,162)(551,161)(552,159)(554,158)(555,157)(556,156)(558,156)(561,155)(563,154)(566,154)(572,153)(577,152)(583,152)(588,151)(594,151)(599,150)(605,149)(610,149)(616,148)(621,148)(627,147)(630,147)(632,147)(634,147)(635,147)(637,147)
\thicklines \dashline[-10]{25}(637,147)(638,147)(639,147)(639,147)(639,147)(640,147)(640,147)(640,147)(640,147)(640,147)(641,147)(641,147)(641,147)(641,147)(641,147)(641,147)(642,147)(642,147)(642,147)(642,147)(643,147)(643,147)(644,147)(645,147)(646,147)(646,147)(648,147)(649,147)(650,147)(652,148)(653,148)(655,149)(656,150)(657,151)(658,152)(659,153)(659,155)(660,156)(661,160)(662,163)(663,166)(664,171)(664,176)(665,184)(665,190)(666,196)(666,212)(667,223)(667,238)(667,256)
\thicklines \dashline[-10]{25}(668,280)(668,313)(668,335)(668,360)(669,390)(669,428)(669,474)(669,533)(669,609)(669,710)(670,846)(670,877)
\thicklines \dashline[-10]{25}(672,812)(672,710)(672,629)(672,564)(672,511)(672,430)(672,372)(672,330)(672,298)(673,273)(673,253)(673,238)(673,225)(673,205)(673,192)(673,182)(674,174)(674,168)(674,164)(674,160)(675,157)(675,155)(675,153)(675,150)(676,147)(676,146)(677,144)(677,143)(678,143)(678,142)(678,141)(679,141)(679,141)(680,140)(680,140)(681,140)(681,140)(682,140)(682,140)(682,140)(682,140)(683,140)(683,140)(683,139)(683,139)(683,139)(683,139)(683,139)(683,139)(683,139)
\thicklines \dashline[-10]{25}(683,139)(683,139)(684,139)(684,139)(684,139)(684,139)(684,139)(684,139)(684,139)(684,139)(684,139)(684,139)(684,139)(684,139)(684,139)(684,139)(684,140)(685,140)(685,140)(685,140)(685,140)(686,140)(686,140)(687,140)(687,141)(688,141)(688,142)(688,142)(689,144)(690,146)(691,149)(692,151)
\thicklines \dashline[-10]{25}(692,142)(692,142)(692,142)(692,142)(692,142)(692,141)(693,141)(693,141)(693,141)(694,140)(695,139)(696,139)(696,138)(697,138)(698,137)(699,137)(701,137)(704,136)(706,136)(709,135)(712,135)(715,134)(718,134)(721,134)(724,134)(727,133)(730,133)(733,133)(736,132)(737,132)(739,132)(740,132)(741,132)(741,132)(741,132)(742,132)(742,132)(742,132)(742,132)(742,132)(742,132)(743,132)(743,132)(743,132)(743,132)(743,132)(743,132)(743,132)(743,132)(743,132)
\thicklines \dashline[-10]{25}(743,132)(744,132)(744,132)(744,132)(744,132)(745,132)(745,132)(746,132)(747,132)(747,132)(748,132)(749,133)(750,133)(750,133)(752,133)(753,134)(755,135)(756,136)(759,139)(762,141)
\thicklines \dashline[-10]{25}(762,132)(767,130)(769,129)(771,128)(774,128)(776,127)(781,127)(790,125)(799,124)(808,123)(818,122)(827,121)(836,120)(846,120)(855,119)(864,118)(873,117)(883,117)(892,116)(901,115)(911,115)(920,114)(929,114)(934,114)(936,113)(939,113)(940,113)(940,113)(941,113)(941,113)(941,113)
\end{picture}
        \end{center}
        \end{minipage}
        \caption{Left : differential BR for $e^+ \, e^-$
	with (thick curve) and without (thin curve) resonances for 
	$(\gamma,x) = ({48.6}^o,0.778)$. 
	The upper, middle and lower dashed curves show the SD contribution 
	for same $x$ and $\gamma = 0^o, {90}^o, {-90}^o$.
	Right :  same with the left for $\gamma = {48.6}^o$ and 
	$x = 1.368$ (solid curve) and $0.631$ (dashed curve).}
        \label{fig:dbr}
\end{figure}
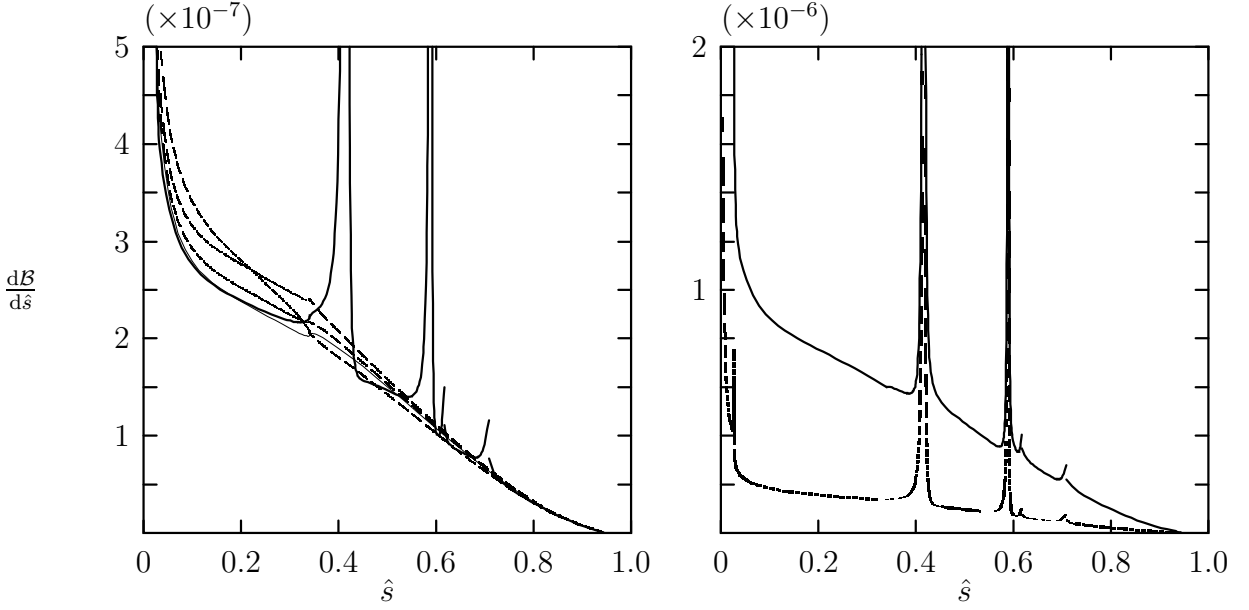	

Now, let me consider the dilepton invariant mass distribution 
of the differential branching-ratio (BR). It can be obtained by
integrating Eq. (\ref{eq:dddr}) over the whole region of variable $z$,  
\begin{equation}
 	\frac{{\rm d}{\cal B}(\s)}{{\rm d}\s} = 
	\int_{-1}^1 {\rm d}z \, 
	\frac{{\rm d}^2{\cal B}(\s,z)}{{\rm d}\s {\rm d}z}
	\, .
	\label{eq:ddbr}
\end{equation}
This gives 
\begin{eqnarray}
	\frac{{\rm d}{\cal B}(\s)}{{\rm d}\s} & = & 
	\frac{4}{3} \, {\cal B}_o \sqrt{1 - \frac{4 \, \ml^2}{\s}} \, \u(\s) \, \left\{ 
	6 \, \left[ \left| {C_9}^{\rm eff} \right|^2 - \left| C_{10} \right|^2 \right] 
		\, \ml^2 \, \left[1 - \s + \mq^2 \right]
	\right. 
	\nonumber \\
	& & \left. 
	+ \left[ \left| {C_9}^{\rm eff} \right|^2 + \left| C_{10} \right|^2 \right]
		\left[ (1 - \mq^2)^2 + \s \, (1 + \mq^2) - 
		2 \, \s^2 + {\u(\s)}^2 \frac{2 \, \ml^2}{\s} \right]
	\right. 
	\nonumber \\
	& & \left. 
	+ 4 \, \left| {C_7}^{\rm eff} \right|^2 \frac{1 + {2\, \ml^2}/\s}{\s}
	\right. 
	\nonumber \\
	& & \left. \; \; \; \;
	 	\times \left[ 2 \, (1 + \mq^2)(1 - \mq^2)^2 - 
		(1 + 14 \, \mq^2 + \mq^4) \, \s - (1 + \mq^2) \, \s^2 \right]
	\right. 
	\nonumber \\
	& & \left. 
	+ 12 \, {\rm Re} \left( {C_9}^{\rm eff} \right)^\ast \, {C_7}^{\rm eff} \,
		\left[ 1 + \frac{2 \, \ml^2}{\s} \right] \, 
		\left[ (1 - \mq^2)^2 - (1 + \mq^2) \, \s \right]
	\right\} \; .
	\label{eq:dbr}
\end{eqnarray}

The distribution of differential BR on dilepton invariant mass 
for $B \rightarrow X_d \, e^+ \, e^-$ is given in Fig. \ref{fig:dbr} 
for various values of $\gamma$ and $x$. 
High sensitivity on $x$ is mostly coming from the CKM factor
in Eq. (\ref{eq:bo}), while the sensitivity on $\gamma$ 
seems significant only in the low $\s$ region. Then, in 
the high $\s$ region the differential BR may be a good test for 
$x$ and makes good the loss of $B^0_d-\bar{B}^0_d$ mixing due to 
the theoretical uncertainties in the treatment of hadron matrix element
$\left< B \left| {\cal O}^\dagger {\cal O} \right| B \right>$. 
Anyway, I have checked that the momentum dependence of 
resonances in the channel is not as large as the result 
in \cite{ldeq}. There is only $\sim 4\%$ reducement 
compared with using $\fv(\mv^2)$ for all region. This 
is because the $1/\s$ suppression in Eq. (\ref{eq:fs}) \cite{nldeq}.

Next, I am going on examining some measurements that are sensitive on
$\gamma$ and less sensitive on $x$. This may be achieved 
by considering the asymmetries and normalizing them with 
the differential BR to eliminate the CKM factor.

\subsection{\bf Forward-backward asymmetry}
\label{sec:fba}

\begin{figure}[t]
        \begin{minipage}[t]{\minitwocolumn}
        \begin{center}
\setlength{\unitlength}{0.240900pt}
\begin{picture}(1049,900)(0,0)
\thicklines \path(220,113)(240,113)
\thicklines \path(985,113)(965,113)
\put(198,113){\makebox(0,0)[r]{$-0.2$}}
\thicklines \path(220,222)(240,222)
\thicklines \path(985,222)(965,222)
\put(198,222){\makebox(0,0)[r]{$-0.1$}}
\thicklines \path(220,331)(240,331)
\thicklines \path(985,331)(965,331)
\put(198,331){\makebox(0,0)[r]{$0$}}
\thicklines \path(220,440)(240,440)
\thicklines \path(985,440)(965,440)
\put(198,440){\makebox(0,0)[r]{$0.1$}}
\thicklines \path(220,550)(240,550)
\thicklines \path(985,550)(965,550)
\put(198,550){\makebox(0,0)[r]{$0.2$}}
\thicklines \path(220,659)(240,659)
\thicklines \path(985,659)(965,659)
\put(198,659){\makebox(0,0)[r]{$0.3$}}
\thicklines \path(220,768)(240,768)
\thicklines \path(985,768)(965,768)
\put(198,768){\makebox(0,0)[r]{$0.4$}}
\thicklines \path(220,877)(240,877)
\thicklines \path(985,877)(965,877)
\put(198,877){\makebox(0,0)[r]{$0.5$}}
\thicklines \path(220,113)(220,133)
\thicklines \path(220,877)(220,857)
\put(220,68){\makebox(0,0){$0$}}
\thicklines \path(373,113)(373,133)
\thicklines \path(373,877)(373,857)
\put(373,68){\makebox(0,0){$0.2$}}
\thicklines \path(526,113)(526,133)
\thicklines \path(526,877)(526,857)
\put(526,68){\makebox(0,0){$0.4$}}
\thicklines \path(679,113)(679,133)
\thicklines \path(679,877)(679,857)
\put(679,68){\makebox(0,0){$0.6$}}
\thicklines \path(832,113)(832,133)
\thicklines \path(832,877)(832,857)
\put(832,68){\makebox(0,0){$0.8$}}
\thicklines \path(985,113)(985,133)
\thicklines \path(985,877)(985,857)
\put(985,68){\makebox(0,0){$1.0$}}
\thicklines \path(220,113)(985,113)(985,877)(220,877)(220,113)
\put(0,495){\makebox(0,0)[l]{\shortstack{$\bar{\cal A}_{\rm FB}$}}}
\put(602,23){\makebox(0,0){$\hat{s}$}}
\thinlines \path(220,331)(220,331)(220,331)(220,331)(220,331)(220,331)(220,331)(220,331)(220,331)(220,331)(220,331)(220,331)(220,331)(220,331)(220,331)(220,331)(220,331)(220,331)(220,331)(220,331)(220,331)(220,331)(220,331)(220,331)(220,331)(220,331)(220,331)(220,331)(220,331)(220,331)
\thinlines \path(220,331)(220,331)(223,294)(225,264)(228,240)(231,221)(234,206)(236,195)(239,186)(242,179)(244,174)(246,172)(247,170)(248,169)(250,168)(251,167)(251,167)(252,167)(252,167)(253,167)(253,167)(253,167)(254,167)(254,167)(254,167)(255,167)(255,167)(255,167)(256,167)(256,167)(257,167)(257,167)(258,167)(259,168)(261,169)(263,171)(269,176)(274,183)(285,201)(296,221)(307,242)(318,264)(328,286)(339,308)(350,329)(361,350)(372,370)(383,390)(394,409)(404,427)(415,444)
\thinlines \path(415,444)(426,461)(437,478)(448,494)(459,509)(469,525)(480,544)
\thinlines \path(480,545)(480,545)(481,548)(482,550)(483,553)(486,558)(488,562)(491,566)(501,580)(522,604)(543,626)(564,645)(585,663)(606,679)(628,694)(649,708)(670,720)(691,732)(712,744)(733,754)(754,763)(775,772)(796,779)(806,782)(817,785)(827,787)(833,788)(838,789)(840,789)(843,789)(844,789)(846,789)(847,790)(848,790)(848,790)(849,790)(850,790)(850,790)(851,790)(852,790)(852,790)(853,790)(854,790)(854,790)(855,790)(856,789)(858,789)(859,789)(861,789)(864,789)(867,788)
\thinlines \path(867,788)(869,787)(875,786)(880,784)(885,780)(890,776)(896,771)(901,764)(906,755)(911,742)(917,726)(919,716)(922,703)(925,689)(927,671)(930,649)(932,621)(935,584)(936,561)(938,533)(938,516)(939,496)(940,472)(940,443)(941,399)
\thicklines \path(220,331)(220,331)(220,331)(220,331)(220,331)(220,331)(220,331)(220,331)(220,331)(220,331)(220,331)(220,331)(220,331)(220,331)(220,331)(220,331)(220,331)(220,331)(220,331)(220,331)(220,331)(220,331)(220,331)(220,331)(220,331)(220,331)(220,331)(220,331)(220,331)(220,331)
\thicklines \path(220,331)(220,331)(221,319)(222,308)(222,297)(223,287)(224,278)(225,270)(226,262)(227,255)(227,248)(228,242)(229,237)(230,231)(231,227)(231,222)(232,218)(233,214)(234,211)(235,207)(236,204)(236,200)(237,195)(238,189)(238,186)(239,185)(239,184)(239,183)(239,183)(239,183)(239,182)(239,182)(239,182)(239,182)(239,182)(239,182)(239,182)(239,182)(239,182)(239,182)(239,182)(239,183)(239,183)(239,183)(239,184)(240,184)(240,185)(240,185)(240,186)(240,186)
\thicklines \path(240,183)(240,183)(240,183)(240,184)(240,185)(240,186)(240,187)(240,188)(240,189)(240,190)(240,192)(240,194)(240,196)(240,198)(240,200)(240,203)(240,206)(240,208)(240,211)(240,214)(240,217)(240,219)(240,220)(240,221)(240,221)(240,221)(240,222)(240,222)(240,222)(240,222)(240,222)(240,222)(240,222)(240,222)(240,222)(240,222)(240,222)(240,222)(240,222)(240,222)(240,222)(240,222)(240,222)(240,222)(240,222)(240,222)(240,222)(240,221)
\thicklines \path(240,206)(240,206)(241,164)(241,150)(241,146)(242,144)(242,143)(242,142)(242,142)(243,142)(243,142)(243,142)(244,143)(245,144)(250,149)(255,153)(260,158)(270,171)(280,188)(290,207)(300,227)(310,248)(320,269)(330,290)(340,311)(350,331)(360,351)(370,370)(380,389)(390,407)(400,425)(410,442)(420,459)(430,476)(440,492)(450,508)(460,525)(470,541)(480,559)
\thicklines \path(480,559)(480,559)(480,560)(480,561)(481,562)(481,563)(482,564)(483,567)(485,572)(488,577)(490,581)(492,585)(495,589)(497,593)(500,597)(502,600)(505,603)(507,606)(509,609)(511,610)(512,611)(512,611)(513,612)(513,612)(514,612)(514,612)(514,612)(514,612)(515,612)(515,612)(515,612)(515,612)(515,612)(515,612)(515,612)(515,612)(515,612)(515,612)(515,612)(515,612)(516,612)(516,612)(516,612)(516,612)(516,612)(517,612)(517,611)(518,611)(518,610)(519,610)(520,609)
\thicklines \path(520,609)(520,608)(521,605)(523,601)(524,595)(525,588)(526,580)(528,568)(529,555)(531,517)(534,464)(536,398)(538,332)
\thicklines \path(538,332)(538,332)(539,302)(540,276)(540,265)(541,255)(541,247)(541,243)(541,240)(541,238)(542,236)(542,234)(542,233)(542,233)(542,233)(542,233)(543,234)(543,235)(543,237)(543,241)(544,247)(544,254)(547,332)(549,408)(552,465)(555,505)(558,534)(561,556)(563,573)(566,586)(572,607)(577,622)(583,634)(588,645)(594,654)(599,663)(605,671)(610,678)(616,686)(621,693)(627,700)(632,707)(638,714)(641,717)(643,720)(645,722)(646,723)(647,723)(648,724)(648,724)(649,724)
\thicklines \path(649,724)(649,724)(649,724)(650,724)(650,725)(650,725)(650,725)(650,725)(651,725)(651,725)(651,725)(651,724)(651,724)(651,724)(652,724)(652,724)(652,724)(653,723)(654,722)(655,721)(655,720)(656,718)(657,716)(657,713)(658,710)(659,706)(659,701)(660,695)(661,679)(663,658)(664,628)(666,588)(667,535)(668,472)(671,334)
\thicklines \path(671,337)(671,337)(672,294)(672,276)(672,261)(672,255)(672,250)(672,246)(672,244)(672,242)(672,241)(673,240)(673,240)(673,239)(673,239)(673,239)(673,239)(673,239)(673,238)(673,238)(673,239)(673,239)(673,239)(673,239)(673,240)(673,241)(673,243)(673,246)(673,250)(674,272)(675,334)(675,401)(676,459)(677,507)(678,546)(679,577)(680,603)(681,624)(682,642)(682,658)(683,672)(684,684)(685,696)(686,708)(687,718)(688,728)(688,733)(688,737)(689,739)(689,740)(689,741)
\thicklines \path(689,741)(689,742)(689,742)(689,742)(689,742)(689,742)(689,742)(690,742)(690,742)(690,742)(690,742)(690,742)(690,742)(690,742)(690,742)(690,742)(690,742)(690,742)(690,742)(690,742)(690,741)(690,741)(690,739)(691,737)(691,734)(691,730)(691,721)(692,709)
\thicklines \path(692,707)(692,707)(693,698)(693,693)(693,689)(694,686)(694,685)(694,684)(694,683)(694,682)(695,681)(695,681)(695,681)(695,681)(695,681)(695,681)(695,681)(695,681)(695,681)(695,681)(696,681)(696,681)(696,681)(696,681)(696,682)(697,683)(698,686)(701,695)(704,703)(706,709)(709,715)(712,721)(715,726)(718,730)(721,735)(724,739)(727,743)(730,748)(733,752)(736,757)(739,762)(741,767)(744,773)(747,779)(749,782)(750,784)(752,787)(752,787)(752,787)(753,788)(753,788)
\thicklines \path(753,788)(753,788)(753,788)(753,788)(754,788)(754,788)(754,788)(754,788)(754,788)(754,788)(754,788)(754,788)(754,788)(754,788)(754,788)(755,788)(755,788)(755,788)(755,787)(756,787)(756,786)(757,784)(758,781)(758,777)(759,772)(760,766)(760,759)(762,739)
\thicklines \path(762,767)(762,767)(764,755)(765,750)(766,747)(767,745)(767,744)(768,743)(768,742)(769,741)(769,740)(769,740)(770,740)(770,740)(770,740)(771,740)(771,740)(771,740)(781,746)(790,754)(799,760)(808,765)(818,769)(827,773)(836,775)(841,776)(846,777)(850,778)(853,778)(854,778)(855,778)(855,778)(856,778)(857,778)(857,778)(857,778)(857,778)(858,778)(858,778)(858,778)(859,778)(859,778)(859,778)(860,778)(860,778)(860,778)(860,778)(861,778)(862,778)(862,778)(863,778)
\thicklines \path(863,778)(864,778)(866,778)(869,777)(871,777)(873,776)(878,775)(883,773)(887,770)(892,766)(897,762)(901,756)(906,748)(911,738)(915,724)(920,707)(925,683)(927,668)(929,649)(932,627)(934,599)(936,561)(937,537)(939,508)(940,469)(940,444)(941,428)(941,408)(941,380)
\end{picture}
        \end{center}
        \end{minipage}
        \hfill
        \begin{minipage}[t]{\minitwocolumn}
        \begin{center}
\setlength{\unitlength}{0.240900pt}
\begin{picture}(1049,900)(0,0)
\thicklines \path(220,113)(240,113)
\thicklines \path(985,113)(965,113)
\put(198,113){\makebox(0,0)[r]{$-0.2$}}
\thicklines \path(220,222)(240,222)
\thicklines \path(985,222)(965,222)
\put(198,222){\makebox(0,0)[r]{$-0.1$}}
\thicklines \path(220,331)(240,331)
\thicklines \path(985,331)(965,331)
\put(198,331){\makebox(0,0)[r]{$0$}}
\thicklines \path(220,440)(240,440)
\thicklines \path(985,440)(965,440)
\put(198,440){\makebox(0,0)[r]{$0.1$}}
\thicklines \path(220,550)(240,550)
\thicklines \path(985,550)(965,550)
\put(198,550){\makebox(0,0)[r]{$0.2$}}
\thicklines \path(220,659)(240,659)
\thicklines \path(985,659)(965,659)
\put(198,659){\makebox(0,0)[r]{$0.3$}}
\thicklines \path(220,768)(240,768)
\thicklines \path(985,768)(965,768)
\put(198,768){\makebox(0,0)[r]{$0.4$}}
\thicklines \path(220,877)(240,877)
\thicklines \path(985,877)(965,877)
\put(198,877){\makebox(0,0)[r]{$0.5$}}
\thicklines \path(220,113)(220,133)
\thicklines \path(220,877)(220,857)
\put(220,68){\makebox(0,0){$0$}}
\thicklines \path(373,113)(373,133)
\thicklines \path(373,877)(373,857)
\put(373,68){\makebox(0,0){$0.2$}}
\thicklines \path(526,113)(526,133)
\thicklines \path(526,877)(526,857)
\put(526,68){\makebox(0,0){$0.4$}}
\thicklines \path(679,113)(679,133)
\thicklines \path(679,877)(679,857)
\put(679,68){\makebox(0,0){$0.6$}}
\thicklines \path(832,113)(832,133)
\thicklines \path(832,877)(832,857)
\put(832,68){\makebox(0,0){$0.8$}}
\thicklines \path(985,113)(985,133)
\thicklines \path(985,877)(985,857)
\put(985,68){\makebox(0,0){$1.0$}}
\thicklines \path(220,113)(985,113)(985,877)(220,877)(220,113)
\put(602,23){\makebox(0,0){$\hat{s}$}}
\thinlines \dashline[-10]{25}(220,331)(220,331)(220,331)(220,331)(220,331)(220,331)(220,331)(220,331)(220,331)(220,331)(220,331)(220,331)(220,331)(220,331)(220,331)(220,331)(220,331)(220,331)(220,331)(220,331)(220,331)(220,331)(220,331)(220,331)(220,331)(
220,331)(220,331)(220,331)(220,331)
\thinlines \dashline[-10]{25}(220,331)(221,319)(222,309)(222,301)(223,293)(224,286)(225,281)(226,276)(227,272)(227,269)(228,268)(228,267)(229,266)(229,265)(229,265)(229,265)(230,265)(230,265)(230,264)(230,264)(230,264)(230,264)(230,264)(230,264)(230,264)(
230,264)(230,264)(230,264)(230,264)(230,264)(230,264)(230,264)(230,264)(230,264)(230,264)(230,264)(230,264)(231,264)(231,264)(231,265)(231,265)(231,265)(231,265)(231,265)(232,266)(232,267)(233,269)(234,272)(235,276)(236,280)
\thinlines \dashline[-10]{25}(236,284)(237,288)(238,291)(238,292)(239,292)(239,293)(239,293)(239,294)(239,294)(239,295)(239,296)(239,297)(239,298)(240,298)(240,301)
\thinlines \dashline[-10]{25}(240,295)(240,296)(240,297)(240,298)(240,299)(240,300)(240,301)(240,302)(240,303)(240,304)(240,305)(240,307)(240,308)(240,309)(240,311)(240,312)(240,313)(240,315)(240,315)(240,316)(240,316)(240,316)(240,316)(240,317)(240,317)(
240,317)(240,317)(240,317)(240,317)(240,317)(240,317)(240,317)(240,317)(240,317)(240,317)(240,317)(240,317)(240,317)(240,317)(240,317)(240,317)(240,317)(240,317)(240,317)(240,317)(240,317)(240,317)(240,317)(240,316)(240,316)
\thinlines \dashline[-10]{25}(240,316)(240,316)(240,315)(240,314)
\thinlines \dashline[-10]{25}(240,310)(241,273)(241,251)(241,241)(242,234)(243,209)(243,198)(244,189)(245,181)(245,174)(246,170)(247,166)(247,163)(247,163)(248,162)(248,161)(248,161)(249,160)(249,160)(249,160)(250,160)(250,160)(250,160)(251,160)(251,161)(
252,161)(253,163)(255,167)(260,177)(270,197)(280,219)(290,240)(300,260)(310,281)(320,300)(330,319)(340,337)(350,354)(360,371)(370,386)(380,401)(390,414)(400,427)(410,438)(420,449)(430,458)(440,465)(445,469)(450,471)(453,472)
\thinlines \dashline[-10]{25}(455,473)(457,473)(458,473)(458,473)(459,473)(460,473)(460,473)(460,474)(461,474)(461,474)(461,474)(462,474)(462,474)(462,474)(462,474)(463,474)(463,474)(463,474)(464,473)(465,473)(465,473)(467,473)(468,473)(469,472)(470,471)(
472,471)(473,470)(475,467)(477,465)(478,463)(478,462)(479,460)(479,459)(480,458)(480,457)(480,453)
\thinlines \dashline[-10]{25}(480,453)(480,453)(480,453)(481,453)(481,453)(481,453)(481,453)(481,453)(481,453)(481,453)(481,453)(481,453)(481,453)(481,453)(481,453)(482,453)(482,453)(482,453)(482,453)(482,453)(482,453)(482,453)(483,453)(483,453)(484,453)(
485,452)(486,452)(488,451)(490,448)(492,445)(495,440)(497,435)(500,428)(502,419)(505,407)(507,392)(509,374)(512,350)(514,320)(517,280)(519,228)(521,163)(523,113)
\thinlines \dashline[-10]{25}(536,201)(538,331)
\thinlines \dashline[-10]{25}(538,331)(540,432)(541,508)(542,537)(543,562)(543,582)(544,599)(545,613)(545,624)(546,634)(547,642)(547,648)(548,653)(549,657)(549,661)(550,664)(551,666)(552,668)(552,669)(553,670)(554,671)(554,672)(555,672)(555,672)(555,672)(
556,672)(556,673)(556,673)(556,673)(557,673)(557,673)(557,673)(557,673)(557,673)(557,673)(558,673)(558,673)(558,673)(558,673)(558,673)(559,672)(560,672)(561,672)(566,670)(572,667)(577,666)(583,664)(588,663)(594,662)(599,661)
\thinlines \dashline[-10]{25}(605,660)(608,659)(610,658)(613,657)(616,656)(619,654)(621,652)(624,650)(627,647)(630,643)(632,638)(635,631)(638,623)(641,612)(643,598)(646,578)(648,566)(649,551)(650,533)(652,511)(653,484)(655,451)(656,409)(657,356)(660,206)(
661,113)
\thinlines \dashline[-10]{25}(671,331)
\thinlines \dashline[-10]{25}(671,331)(672,412)(672,479)(672,535)(673,579)(673,615)(674,644)(674,667)(675,685)(675,700)(675,711)(676,721)(676,728)(677,734)(677,739)(678,743)(678,746)(678,748)(679,750)(679,751)(680,752)(680,752)(680,753)(680,753)(680,753)(
681,754)(681,754)(681,754)(681,754)(681,754)(681,754)(681,754)(681,754)(681,754)(681,754)(681,754)(681,754)(681,754)(681,754)(681,754)(681,754)(681,754)(681,754)(681,754)(682,754)(682,754)(682,754)(682,754)(682,754)(682,754)
\thinlines \dashline[-10]{25}(682,754)(682,753)(683,753)(683,752)(684,751)(685,749)(686,746)(687,742)(688,738)(688,732)(689,730)(689,727)(690,725)(690,724)(690,724)(690,724)(690,723)(690,723)(690,723)(690,723)(690,723)(690,723)(690,723)(690,723)(690,723)(
690,723)(690,723)(690,723)(690,723)(691,723)(691,723)(691,724)(691,724)(691,724)(691,725)(691,725)(692,731)
\thinlines \dashline[-10]{25}(692,756)(692,758)(693,759)(693,760)(693,760)(693,761)(693,761)(694,761)(694,761)(694,761)(694,761)(694,762)(694,762)(694,762)(694,762)(694,762)(694,762)(694,762)(695,762)(695,761)(695,761)(695,761)(696,761)(696,760)(698,759)(
701,756)(704,754)(706,753)(709,752)(712,751)(715,750)(718,749)(721,748)(724,747)(727,746)(730,745)(733,743)(736,741)(739,738)(741,734)(744,728)(747,720)(750,708)(753,692)(755,684)(755,680)(756,679)(756,677)(756,676)(757,676)
\thinlines \dashline[-10]{25}(757,676)(757,675)(757,675)(757,675)(757,675)(757,675)(757,675)(757,675)(758,675)(758,675)(758,675)(758,676)(758,676)(758,677)(759,678)(759,680)(760,686)(760,694)(762,706)
\thinlines \dashline[-10]{25}(762,771)(763,777)(764,780)(764,782)(765,784)(765,786)(766,787)(767,788)(767,789)(767,789)(768,789)(768,789)(768,790)(769,790)(769,790)(769,790)(769,790)(770,790)(770,790)(771,790)(771,789)(776,788)(778,787)(781,786)(782,786)(
783,786)(784,786)(785,786)(786,786)(787,786)(788,785)(789,785)(789,785)(790,785)(790,785)(790,785)(790,785)(791,785)(791,785)(791,785)(792,785)(792,785)(792,785)(793,785)(793,785)(793,785)(794,785)(796,786)(797,786)(799,786)
\thinlines \dashline[-10]{25}(804,786)(808,787)(818,788)(827,788)(832,789)(834,789)(836,789)(837,789)(839,789)(840,789)(841,789)(842,789)(842,789)(842,789)(843,789)(843,789)(843,789)(844,789)(844,789)(844,789)(844,789)(845,789)(845,789)(845,789)(846,789)(
846,789)(847,789)(847,789)(848,789)(849,789)(850,789)(853,789)(855,789)(857,789)(860,788)(864,787)(869,786)(873,785)(878,783)(883,780)(887,777)(892,773)(897,768)(901,761)(906,753)(911,742)(915,729)(920,711)(925,686)(927,671)
\thinlines \dashline[-10]{25}(929,652)(932,629)(934,601)(936,563)(937,539)(939,509)(940,470)(940,444)(941,428)(941,408)(941,380)
\thicklines \path(220,331)(220,331)(220,331)(220,331)(220,331)(220,331)(220,331)(220,331)(220,331)(220,331)(220,331)(220,331)(220,331)(220,331)(220,331)(220,331)(220,331)(220,331)(220,331)(220,331)(220,331)(220,331)(220,331)(220,331)(220,331)(220,331)(220
,331)(220,331)(220,331)(220,331)
\thicklines \path(220,331)(220,331)(221,319)(222,308)(222,297)(223,287)(224,278)(225,270)(226,262)(227,255)(227,249)(228,243)(229,237)(230,232)(231,227)(231,223)(232,219)(233,215)(234,211)(235,208)(236,204)(236,200)(237,195)(238,190)(238,187)(239,184)(239
,183)(239,182)(239,182)(239,181)(239,181)(239,181)(239,181)(239,181)(239,181)(240,181)(240,180)(240,180)
\thicklines \path(240,182)(240,182)(240,182)(240,182)(240,182)(240,183)(240,183)(240,183)(240,183)(240,183)(240,184)(240,184)(240,184)(240,185)(240,185)(240,185)(240,185)(240,185)(240,185)(240,185)(240,185)(240,185)(240,185)(240,185)(240,185)(240,185)(240
,185)(240,185)(240,185)(240,185)(240,185)(240,185)(240,185)(240,185)(240,185)(240,185)(240,185)(240,185)(240,185)(240,185)(240,185)(240,185)(240,185)(240,185)(240,185)(240,185)(240,185)(240,184)(240,184)(240,183)(240,183)
\thicklines \path(240,183)(240,182)(240,179)(240,177)(240,173)(240,169)
\thicklines \path(240,165)(240,165)(241,150)(241,154)(241,156)(242,156)(243,157)(245,160)(250,164)(255,168)(260,173)(270,187)(280,205)(290,225)(300,246)(310,268)(320,290)(330,311)(340,332)(350,353)(360,373)(370,392)(380,411)(390,429)(400,447)(410,464)(420
,480)(430,497)(440,513)(450,529)(460,545)(470,561)(480,580)
\thicklines \path(480,580)(480,580)(483,585)(485,589)(488,593)(490,597)(492,600)(495,604)(497,608)(500,612)(502,616)(505,620)(507,624)(509,628)(512,632)(514,636)(517,639)(518,641)(519,642)(520,643)(520,643)(521,644)(521,644)(521,644)(522,644)(522,644)(522
,644)(522,644)(522,644)(522,644)(522,644)(522,644)(522,644)(523,644)(523,644)(523,644)(523,644)(523,644)(523,644)(523,644)(523,644)(523,644)(524,644)(524,644)(524,644)(525,643)(525,643)(526,641)(526,640)(527,638)(528,635)
\thicklines \path(528,635)(528,632)(529,629)(529,624)(530,618)(531,604)(532,583)(534,554)(535,515)(536,464)(538,332)
\thicklines \path(538,332)(538,332)(539,257)(540,184)(540,151)(541,123)(541,113)
\thicklines \path(543,113)(543,115)(543,135)(544,179)(545,268)(547,342)(548,398)(549,441)(551,473)(552,498)(554,518)(555,534)(558,559)(561,577)(563,591)(566,601)(572,618)(577,631)(583,642)(588,651)(594,659)(599,667)(605,674)(610,681)(616,688)(621,695)(627
,702)(632,709)(638,717)(643,725)(649,733)(652,737)(653,739)(655,741)(655,741)(656,741)(656,742)(656,742)(656,742)(657,742)(657,742)(657,742)(657,742)(657,742)(657,742)(658,742)(658,742)(658,742)(658,742)(658,742)(659,741)
\thicklines \path(659,741)(659,741)(659,741)(660,740)(660,739)(661,737)(661,735)(662,731)(663,726)(664,719)(664,711)(665,700)(666,685)(666,667)(667,643)(668,612)(668,574)(670,470)(671,334)
\thicklines \path(671,337)(671,337)(672,226)(672,175)(672,130)(672,113)
\thicklines \path(673,113)(673,124)(674,191)(675,326)(675,382)(675,429)(676,467)(676,499)(677,548)(678,583)(679,610)(680,630)(681,646)(682,660)(682,671)(683,681)(684,690)(685,699)(686,706)(687,714)(688,720)(688,723)(688,726)(689,727)(689,727)(689,728)(689
,728)(689,728)(689,728)(689,728)(689,728)(689,728)(689,728)(689,728)(689,728)(689,728)(689,728)(689,728)(690,728)(690,728)(690,728)(690,728)(690,728)(690,728)(690,728)(690,727)(690,727)(690,726)(690,725)(691,722)(691,716)
\thicklines \path(691,716)(691,708)(692,697)
\thicklines \path(692,701)(692,701)(693,694)(693,691)(693,690)(693,688)(694,688)(694,687)(694,687)(694,686)(694,686)(694,686)(694,686)(694,686)(694,686)(694,686)(695,686)(695,686)(695,686)(695,686)(695,686)(695,686)(696,687)(696,689)(698,694)(701,703)(704
,710)(706,716)(709,721)(712,725)(715,729)(718,733)(721,737)(724,740)(727,744)(730,747)(733,751)(736,755)(739,759)(741,763)(744,767)(747,772)(749,774)(750,776)(752,778)(752,779)(753,779)(753,779)(753,779)(753,779)(754,779)
\thicklines \path(754,779)(754,779)(754,779)(754,779)(754,779)(754,779)(754,779)(754,779)(754,779)(754,779)(754,779)(755,779)(755,779)(755,779)(755,779)(756,778)(756,778)(757,776)(757,775)(758,773)(758,770)(759,765)(760,759)(760,752)(762,734)
\thicklines \path(762,756)(762,756)(763,751)(764,746)(765,743)(766,741)(766,741)(767,740)(767,740)(767,740)(767,740)(768,739)(768,739)(768,739)(769,739)(769,739)(769,739)(769,739)(770,740)(771,740)(771,741)(781,750)(790,758)(799,764)(808,768)(818,772)(827
,775)(832,777)(836,778)(841,779)(846,779)(848,780)(850,780)(851,780)(853,780)(854,780)(854,780)(855,780)(855,780)(856,780)(856,780)(856,780)(857,780)(857,780)(857,780)(857,780)(858,780)(858,780)(858,780)(859,780)(859,780)
\thicklines \path(859,780)(860,780)(860,780)(861,780)(862,780)(863,780)(864,780)(866,780)(869,779)(871,779)(873,778)(878,777)(883,775)(887,772)(892,768)(897,763)(901,757)(906,749)(911,739)(915,726)(920,708)(925,684)(927,669)(929,650)(932,628)(934,599)(936
,562)(937,538)(939,508)(940,470)(940,444)(941,428)(941,408)(941,380)
\end{picture}
        \end{center}
        \end{minipage}
        \caption{Left : FB asymmetry for $e^+ \, e^-$
	with (thick curve) and without (thin curve) resonances
	for $(\gamma,x) = ({48.6}^o,0.778)$.
	Right : same with the left for same $x$ and 
	$\gamma ={90}^o$ (solid line) and $0^o$ (dashed line).}
        \label{fig:dfba}
\end{figure}
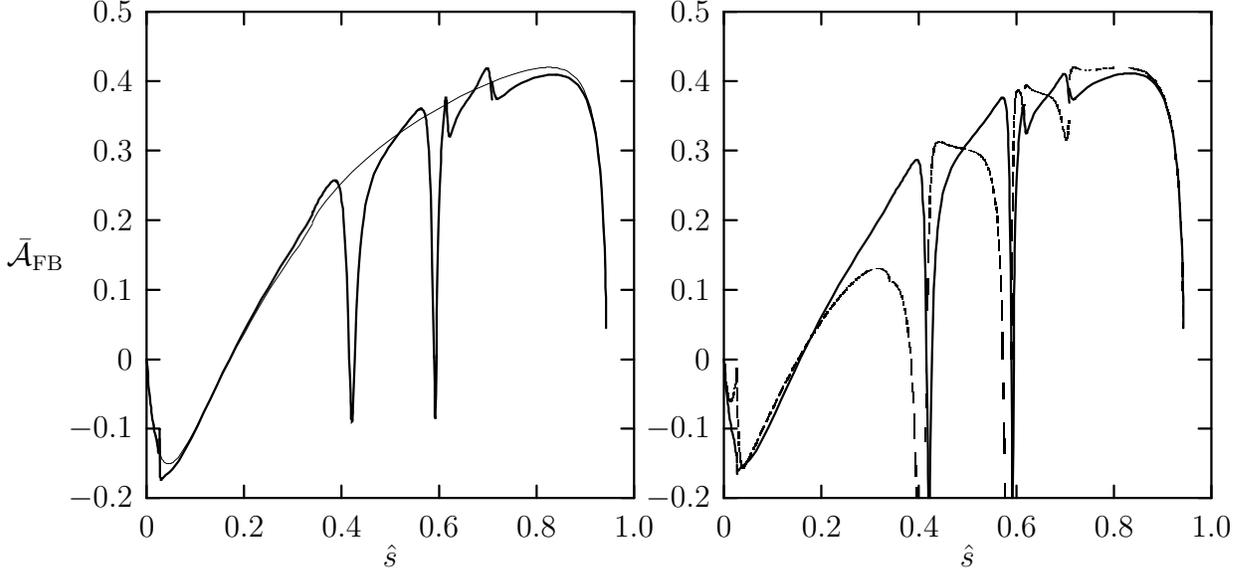

First, I provide the FB asymmetry. 
The normalized FB asymmetry is defined as follows \cite{fba}
\begin{equation}
	\bar{\cal A}_{\rm FB}(\s) = \frac{\displaystyle 
	\int_0^1 {\rm d}z \, 
	\frac{{\rm d}^2{\cal B}(\s,z)}{{\rm d}\s {\rm d}z}
	- \int_{-1}^0 {\rm d}z \, 
	\frac{{\rm d}^2{\cal B}(\s,z)}{{\rm d}\s {\rm d}z}
	}{\displaystyle 
	\int_0^1 {\rm d}z \, 
	\frac{{\rm d}^2{\cal B}(\s,z)}{{\rm d}\s {\rm d}z}
	+ \int_{-1}^0 {\rm d}z \, 
	\frac{{\rm d}^2{\cal B}(\s,z)}{{\rm d}\s {\rm d}z}
	} 
	= \frac{{{\rm d}{\cal A}_{\rm FB}(\s)}/{{\rm d}\s}
	}{{{\rm d}{\cal B}(\s)}/{{\rm d}\s}} \; .
	\label{eq:nfba}
\end{equation}
Then, after integrating Eq. (\ref{eq:dbr}) properly, the nominator reads
\begin{equation}
	\frac{{\rm d}{\cal A}_{\rm FB}(\s)}{{\rm d}\s} = 
	-4 \, {\cal B}_o \sqrt{1 - \frac{4 \, \ml^2}{\s}} \,
	{\u(\s)}^2 \, C_{10} \, \left[  
	{\rm Re} \, \left( {C_9}^{\rm eff} \right)^\ast \, \s 
	+ 2 \, {C_7}^{\rm eff} \, \, \left( 1 + \mq^2 \right)
	\right] \; . 
	\label{eq:dfba}
\end{equation}

In Fig. \ref{fig:dfba}, I plot the FB asymmetry for 
$B \rightarrow X_d \, e^+ \, e^-$ with and without the resonances 
in the left figure, while in the right one with varying $\gamma$ 
and keeping $x$ to be constant.

\subsection{\bf CP asymmetry}
\label{sec:cpa}

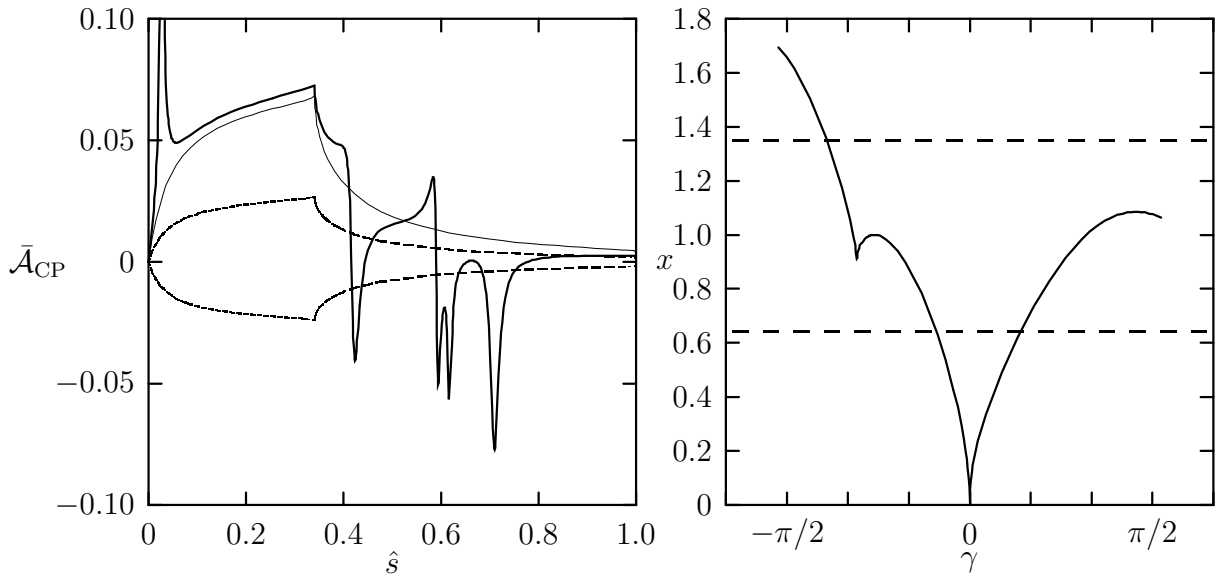
\begin{figure}[t]
        \begin{minipage}[t]{\minitwocolumn}
        \begin{center}
\setlength{\unitlength}{0.240900pt}
\begin{picture}(1049,900)(0,0)
\thicklines \path(220,113)(240,113)
\thicklines \path(985,113)(965,113)
\put(198,113){\makebox(0,0)[r]{$-0.10$}}
\thicklines \path(220,304)(240,304)
\thicklines \path(985,304)(965,304)
\put(198,304){\makebox(0,0)[r]{$-0.05$}}
\thicklines \path(220,495)(240,495)
\thicklines \path(985,495)(965,495)
\put(198,495){\makebox(0,0)[r]{$0$}}
\thicklines \path(220,686)(240,686)
\thicklines \path(985,686)(965,686)
\put(198,686){\makebox(0,0)[r]{$0.05$}}
\thicklines \path(220,877)(240,877)
\thicklines \path(985,877)(965,877)
\put(198,877){\makebox(0,0)[r]{$0.10$}}
\thicklines \path(220,113)(220,133)
\thicklines \path(220,877)(220,857)
\put(220,68){\makebox(0,0){$0$}}
\thicklines \path(373,113)(373,133)
\thicklines \path(373,877)(373,857)
\put(373,68){\makebox(0,0){$0.2$}}
\thicklines \path(526,113)(526,133)
\thicklines \path(526,877)(526,857)
\put(526,68){\makebox(0,0){$0.4$}}
\thicklines \path(679,113)(679,133)
\thicklines \path(679,877)(679,857)
\put(679,68){\makebox(0,0){$0.6$}}
\thicklines \path(832,113)(832,133)
\thicklines \path(832,877)(832,857)
\put(832,68){\makebox(0,0){$0.8$}}
\thicklines \path(985,113)(985,133)
\thicklines \path(985,877)(985,857)
\put(985,68){\makebox(0,0){$1.0$}}
\thicklines \path(220,113)(985,113)(985,877)(220,877)(220,113)
\put(0,495){\makebox(0,0)[l]{\shortstack{$\bar{\cal A}_{\rm CP}$}}}
\put(602,23){\makebox(0,0){$\hat{s}$}}
\thicklines \path(220,495)(220,495)(220,495)(220,495)(220,495)(220,495)(220,495)(220,495)(220,495)(220,495)(220,495)(220,495)(220,495)(220,495)(220,495)(220,495)(220,495)(220,495)(220,495)(220,495)(220,495)(220,495)(220,495)(220,495)(220,495)(220,495)(220
,495)(220,495)(220,495)(220,495)
\thicklines \path(220,495)(220,495)(225,548)(228,577)(229,595)(231,617)(232,646)(233,664)(234,686)(234,713)(235,746)(236,835)(237,877)
\thicklines \path(245,877)(245,848)(246,818)(246,794)(247,775)(248,746)(249,735)(250,726)(251,718)(251,712)(253,703)(254,696)(255,693)(255,691)(256,689)(257,688)(258,685)(259,684)(259,684)(260,683)(261,683)(261,683)(261,683)(262,682)(262,682)(262,682)(263
,682)(263,682)(263,682)(264,682)(264,682)(264,682)(265,682)(265,682)(266,683)(267,683)(269,684)(274,687)(285,694)(296,702)(307,709)(318,715)(328,721)(339,726)(350,731)(361,735)(372,739)(383,742)(394,746)(404,749)(415,752)
\thicklines \path(415,752)(426,756)(437,759)(448,762)(459,765)(469,768)(480,772)
\thicklines \path(480,772)(480,772)(481,756)(482,749)(483,739)(484,732)(486,727)(488,718)(491,710)(496,699)(501,691)(507,686)(509,684)(512,682)(514,681)(517,680)(520,679)(521,679)(522,678)(523,678)(524,677)(524,677)(525,676)(526,675)(526,674)(527,673)(528
,672)(528,670)(529,668)(530,662)(531,658)(532,653)(532,647)(533,639)(534,620)(535,607)(536,592)(538,505)(539,450)(541,399)(541,377)(542,361)(543,349)(543,342)(544,340)(545,342)(545,347)(546,354)(547,371)(549,391)(551,427)
\thicklines \path(551,427)(554,455)(557,476)(559,492)(562,505)(564,514)(570,527)(572,532)(575,536)(580,542)(585,546)(591,549)(596,551)(606,555)(617,558)(622,560)(628,562)(633,564)(638,567)(643,572)(646,575)(649,578)(651,582)(654,587)(656,593)(659,601)(662
,611)(664,622)(665,625)(666,627)(666,628)(667,628)(668,624)(668,617)(669,603)(670,580)(671,497)(672,440)(672,383)(673,338)(674,311)(674,302)(675,305)(676,329)(677,356)(679,380)(680,398)(681,405)(681,411)(682,416)(683,419)
\thicklines \path(683,419)(683,422)(684,423)(685,423)(685,421)(686,417)(687,411)(687,401)(688,388)(689,344)(690,313)(691,279)(696,364)(697,384)(697,400)(698,413)(698,424)(700,440)(701,452)(702,461)(704,467)(706,477)(708,480)(709,483)(712,488)(714,491)(717
,493)(718,494)(720,495)(721,496)(722,496)(723,497)(723,497)(724,497)(725,497)(725,497)(726,497)(727,497)(727,497)(728,497)(729,497)(729,497)(730,497)(731,496)(732,496)(733,496)(734,495)(735,494)(737,493)(738,491)(739,489)
\thicklines \path(739,489)(741,487)(742,484)(743,480)(744,476)(746,470)(748,455)(750,445)(751,431)(754,394)(756,338)(759,267)(760,233)(761,219)(762,209)(762,203)(763,200)(764,203)(764,209)(769,318)(772,366)(773,384)(775,400)(777,423)(780,440)(783,453)(785
,462)(788,469)(790,474)(793,479)(796,482)(801,488)(806,491)(812,494)(817,496)(822,498)(827,499)(833,500)(838,501)(843,502)(848,502)(854,503)(859,503)(869,504)(875,504)(880,504)(885,504)(890,504)(896,504)(901,505)(906,505)
\thicklines \path(906,505)(911,505)(917,505)(922,505)(925,505)(927,505)(930,505)(932,505)(935,505)(938,505)(940,505)(943,505)(953,505)(955,505)(956,505)(959,505)(960,505)(961,505)(963,505)(964,505)(965,505)(967,505)(969,505)(971,505)(972,505)(973,505)(974
,505)(976,505)(977,505)(978,505)(980,505)(980,505)(981,505)(982,505)(982,505)(983,505)(984,505)(984,505)
\thinlines \path(220,495)(220,495)(220,495)(220,495)(220,495)(220,495)(220,495)(220,495)(220,495)(220,495)(220,495)(220,495)(220,495)(220,495)(220,495)(220,495)(220,495)(220,495)(220,495)(220,495)(220,495)(220,495)(220,495)(220,495)(220,495)(220,495)(220,
495)(220,495)(220,495)(220,495)
\thinlines \path(220,495)(220,495)(225,532)(231,560)(242,600)(247,615)(253,627)(263,647)(274,662)(285,674)(296,684)(307,692)(318,699)(328,705)(339,710)(350,715)(361,719)(372,723)(383,726)(394,730)(404,733)(415,736)(426,738)(437,741)(448,744)(459,747)(464,
748)(469,750)(472,751)(475,752)(476,753)(478,754)(479,755)(479,755)(480,755)(480,756)(480,757)
\thinlines \path(480,756)(480,756)(481,737)(482,730)(483,719)(484,710)(486,703)(488,692)(491,683)(496,668)(501,656)(512,637)(522,623)(543,602)(564,586)(585,574)(606,565)(628,557)(649,551)(670,546)(691,541)(712,537)(733,534)(754,531)(775,529)(796,526)(817,
524)(838,522)(859,521)(880,519)(901,518)(922,516)(943,515)(964,514)(969,514)(972,514)(974,513)(977,513)(978,513)(980,513)(980,513)(981,513)(982,512)(982,512)(983,512)(984,511)(984,510)
\thinlines \dashline[-10]{25}(220,495)(220,495)(220,495)(220,495)(220,495)(220,495)(220,495)(220,495)(220,495)(220,495)(220,495)(220,495)(220,495)(220,495)(220,495)(220,495)(220,495)(220,495)(220,495)(220,495)(220,495)(220,495)(220,495)(220,495)(220,495)(
220,495)(220,495)(220,495)(220,495)
\thinlines \dashline[-10]{25}(220,495)(225,510)(231,522)(236,531)(242,538)(253,549)(263,556)(274,562)(285,567)(296,570)(307,574)(318,576)(328,578)(339,580)(350,582)(361,584)(372,585)(383,586)(394,587)(404,589)(415,590)(426,591)(437,592)(448,593)(459,594)(
464,594)(469,595)(472,595)(475,596)(476,596)(478,596)(479,597)(479,597)(480,597)(480,597)(480,598)
\thinlines \dashline[-10]{25}(480,597)(481,590)(482,587)(483,583)(484,580)(486,578)(488,573)(491,570)(496,564)(501,560)(512,552)(522,547)(543,538)(564,532)(585,527)(606,524)(628,521)(649,518)(670,516)(691,514)(712,512)(733,511)(754,510)(775,509)(796,508)(
817,507)(838,506)(859,506)(880,505)(901,504)(922,504)(943,503)(964,503)(969,503)(972,503)(974,503)(977,503)(978,502)(980,502)(980,502)(981,502)(982,502)(982,502)(983,502)(984,502)(984,501)
\thinlines \dashline[-10]{25}(220,495)(220,495)(220,495)(220,495)(220,495)(220,495)(220,495)(220,495)(220,495)(220,495)(220,495)(220,495)(220,495)(220,495)(220,495)(220,495)(220,495)(220,495)(220,495)(220,495)(220,495)(220,495)(220,495)(220,495)(220,495)(
220,495)(220,495)(220,495)(220,495)
\thinlines \dashline[-10]{25}(220,495)(225,480)(231,469)(236,460)(242,454)(253,444)(263,437)(274,432)(285,428)(296,425)(307,422)(318,420)(328,418)(339,417)(350,415)(361,414)(372,413)(383,412)(394,411)(404,410)(415,409)(426,408)(437,407)(448,407)(459,406)(
464,405)(469,405)(472,404)(475,404)(476,404)(478,404)(479,403)(479,403)(480,403)(480,403)(480,403)
\thinlines \dashline[-10]{25}(480,403)(481,408)(482,411)(483,414)(484,417)(486,419)(488,423)(491,426)(501,435)(512,441)(522,446)(543,454)(564,459)(585,464)(606,467)(628,470)(649,473)(670,475)(691,476)(712,478)(733,479)(754,480)(775,481)(796,482)(817,483)(
838,484)(859,484)(880,485)(901,486)(922,486)(943,487)(964,487)(969,487)(972,487)(974,487)(977,488)(978,488)(980,488)(980,488)(981,488)(982,488)(982,488)(983,488)(984,488)(984,489)
\end{picture}
        \end{center}
        \end{minipage}
        \hfill
        \begin{minipage}[t]{\minitwocolumn}
        \begin{center}
\setlength{\unitlength}{0.240900pt}
\begin{picture}(1049,900)(0,0)
\thicklines \path(220,113)(240,113)
\thicklines \path(985,113)(965,113)
\put(198,113){\makebox(0,0)[r]{$0$}}
\thicklines \path(220,198)(240,198)
\thicklines \path(985,198)(965,198)
\put(198,198){\makebox(0,0)[r]{$0.2$}}
\thicklines \path(220,283)(240,283)
\thicklines \path(985,283)(965,283)
\put(198,283){\makebox(0,0)[r]{$0.4$}}
\thicklines \path(220,368)(240,368)
\thicklines \path(985,368)(965,368)
\put(198,368){\makebox(0,0)[r]{$0.6$}}
\thicklines \path(220,453)(240,453)
\thicklines \path(985,453)(965,453)
\put(198,453){\makebox(0,0)[r]{$0.8$}}
\thicklines \path(220,537)(240,537)
\thicklines \path(985,537)(965,537)
\put(198,537){\makebox(0,0)[r]{$1.0$}}
\thicklines \path(220,622)(240,622)
\thicklines \path(985,622)(965,622)
\put(198,622){\makebox(0,0)[r]{$1.2$}}
\thicklines \path(220,707)(240,707)
\thicklines \path(985,707)(965,707)
\put(198,707){\makebox(0,0)[r]{$1.4$}}
\thicklines \path(220,792)(240,792)
\thicklines \path(985,792)(965,792)
\put(198,792){\makebox(0,0)[r]{$1.6$}}
\thicklines \path(220,877)(240,877)
\thicklines \path(985,877)(965,877)
\put(198,877){\makebox(0,0)[r]{$1.8$}}
\thicklines \path(220,113)(220,133)
\thicklines \path(220,877)(220,857)
\thicklines \path(316,113)(316,133)
\thicklines \path(316,877)(316,857)
\put(316,68){\makebox(0,0){$-\pi/2$}}
\thicklines \path(411,113)(411,133)
\thicklines \path(411,877)(411,857)
\thicklines \path(507,113)(507,133)
\thicklines \path(507,877)(507,857)
\thicklines \path(603,113)(603,133)
\thicklines \path(603,877)(603,857)
\put(603,68){\makebox(0,0){$0$}}
\thicklines \path(698,113)(698,133)
\thicklines \path(698,877)(698,857)
\thicklines \path(794,113)(794,133)
\thicklines \path(794,877)(794,857)
\thicklines \path(889,113)(889,133)
\thicklines \path(889,877)(889,857)
\put(889,68){\makebox(0,0){$\pi/2$}}
\thicklines \path(985,113)(985,133)
\thicklines \path(985,877)(985,857)
\thicklines \path(220,113)(985,113)(985,877)(220,877)(220,113)
\multiput(230,385)(48,0){16}{\line(1,0){24}}
\multiput(230,685)(48,0){16}{\line(1,0){24}}
\put(110,495){\makebox(0,0)[l]{\shortstack{$x$}}}
\put(602,23){\makebox(0,0){$\gamma$}}
\thicklines \path(302,832)(302,832)(315,817)(327,799)(352,752)(377,690)(402,611)(415,562)(421,531)(423,521)(423,516)(424,509)(425,500)(426,502)(426,508)(427,511)(429,517)(430,521)(432,524)(434,526)(437,530)(438,532)(440,533)(441,534)(443,535)(444,536)(446,536)(447,537)(448,537)(448,537)(449,537)(450,537)(451,537)(452,537)(452,537)(453,537)(454,537)(455,537)(455,537)(457,537)(459,537)(462,536)(465,534)(471,530)(477,525)(490,512)(502,494)(527,446)(552,381)(565,341)(577,294)
\thicklines \path(577,294)(584,267)(590,236)(593,218)(596,197)(598,185)(599,171)(600,163)(601,153)(602,141)(603,113)(603,139)(604,150)(606,165)(607,176)(609,185)(615,214)(628,256)(653,322)(678,377)(703,425)(728,466)(753,502)(778,530)(790,542)(803,552)(815,560)(822,563)(828,566)(834,569)(840,570)(843,571)(847,572)(850,573)(853,573)(856,573)(858,573)(859,574)(860,574)(861,574)(861,574)(862,574)(863,574)(864,574)(865,574)(865,574)(866,574)(867,574)(868,574)(870,573)(872,573)
\thicklines \path(872,573)(875,573)(878,573)(884,571)(890,570)(903,565)
\end{picture}
        \end{center}
        \end{minipage}
        \caption{Left : CP asymmetry for $e^+ \, e^-$ 
	with (solid thick curve) and without (solid thin curve) 
	resonances for $(\gamma, x) = ({48.6}^o, 0.778)$. 
	The upper and lower dashed curves show 
	the SD contribution for same $x$ and 
	$\gamma = {90}^o, {-90}^o$.
	Right : $x$ as a function of $\gamma$ where $\bar{\cal A}_{\rm CP} = 0$.
	The dashed lines denote the upper and lower bounds of $x$ in the SM.}
        \label{fig:cpa}
\end{figure}

Doing same treatment as \cite{cpvf} in the 
amplitude level, the normalized CP asymmetry can be 
written simply as
\begin{equation}
	\bar{\cal A}_{\rm CP}(\s) = \frac{\displaystyle
	{{\rm d}{\cal B}(\s)}/{{\rm d}\s}
	- 
	{{\rm d}\bar{\cal B}(\s)}/{{\rm d}\s}
	}{\displaystyle 
	{{\rm d}{\cal B}(\s)}/{{\rm d}\s}
	+ 
	{{\rm d}\bar{\cal B}(\s)}/{{\rm d}\s}
	}	
	= \frac{\displaystyle
	-2 \, {{\rm d}{\cal A}_{\rm CP}(\s)}/{{\rm d}\s}
	}{
	{{\rm d}{\cal B}(\s)}/{{\rm d}\s} + 
	2 \, {{\rm d}{\cal A}_{\rm CP}(\s)}/{{\rm d}\s}} \; ,
	\label{eq:ncpa}
\end{equation}
where $\cal B$ and $\bar{\cal B}$ denote the BR of 
$\bar{b} \rightarrow q \, \l^+ \, \l^-$ and its complex conjugate 
$b \rightarrow \bar{q} \, \l^+ \, \l^-$ respectively.
For convenience, ${C_9}^{\rm eff}$ is divided to be two terms
according to the factor ${V_{uq}^\ast \, V_{ub}}/{V_{tq}^\ast \, V_{tb}}$
as below
\begin{equation}
	{C_9}^{\rm eff} = \bar{C}_9 + 
	\frac{V_{uq}^\ast \, V_{ub}}{V_{tq}^\ast \, V_{tb}} \, 
		{C_9}^{\rm CP} \; .
	\label{eq:c9d}
\end{equation}
Then, the result for the differential CP asymmetry is 
\begin{eqnarray}
	\frac{{\rm d}{\cal A}_{\rm CP}(\s)}{{\rm d}\s} & = & 
	\frac{4}{3} {\cal B}_0 \, \sqrt{1 - \frac{4 \, \ml^2}{\s}} \,
	\u(\s) \, {\rm Im} \left[ 
	\frac{V_{uq}^\ast \, V_{ub}}{V_{tq}^\ast \, V_{tb}}\right] \,
	\nonumber \\
	& & \times \left\{  
	{\rm Im} \left[ {\bar{C}_9}^\ast \, {C_9}^{\rm CP} \right] \left[
		(1 - \mq^2)^2 + \s \, (1 + \mq^2) - 
		2 \, \s^2 + {\u(\s)}^2 \frac{2 \, \ml^2}{\s}
	\right. \right.
	\nonumber \\
	& & \left. \left. \; \; \; \; \; \; \; \; \; \; \;
\; \; \; \; \; \; \; \; \;
		+ 6 \, \ml^2 \, (1 - \s + \mq^2) \right]
	\right.
	\nonumber \\
	& & \left.
	+ 6 \, {\rm Im} 
		\left[ {C_7}^{\rm eff} \, {C_9}^{\rm CP} \right] \,
		\left[ 1 + \frac{2 \, \ml^2}{\s} \right] \, 
		\left[ (1 - \mq^2)^2 - (1 + \mq^2) \, \s \right]
	\right\} \; .
	\label{eq:dcpa}
\end{eqnarray}

In Fig. \ref{fig:cpa}, I give the distribution of $\bar{\cal A}_{\rm CP}$
(left figure) with (solid thick line) and 
without (solid thin line) resonances. It is easily understood that in 
the present case, the dependence on $\gamma$ is large, because of the 
appearence of factor $r \sin \gamma$ from the CKM factor 
in Eq. (\ref{eq:dcpa}). Moreover, it is clear that $\bar{\cal A}_{\rm CP}$
will be non-zero if the imaginary part of Eq. (\ref{eq:ckmfactor}) 
is non-zero. Anyway, a condition that $\bar{\cal A}_{\rm CP} = 0 $
for $q = d$ in the SM is satisfied by the following equation, 
\begin{equation}
	x^2 = \sin^2 \gamma \left[ 1 + \frac{1}{4} \left( 
		1 - \sqrt{\left| 3 + 4 \, {\rm cot} \, \gamma \right|} \right)^2 
		\right] \; ,
	\label{eq:xgamma}
\end{equation}
by using Eq. (\ref{eq:r}). The right figure in Fig. \ref{fig:cpa}
is plotted based on this equation. As depicted in the figure, 
there are still allowed regions of $\gamma$ where $\bar{\cal A}_{\rm CP} = 0$. 
Notice again that from Eq. (\ref{eq:ckmfactor}), 
$\bar{\cal A}_{\rm CP}(B \rightarrow X_s \, \l^+ \, \l^-)$
would be $\sim 5\%$ of 
$\bar{\cal A}_{\rm CP}(B \rightarrow X_d \, \l^+ \, \l^-)$ 
due to the suppression of $\lambda^2$.

\subsection{\bf Lepton-polarization asymmetry}
\label{sec:lpa}

\begin{figure}[t]
        \begin{minipage}[t]{\minitwocolumn}
        \begin{center}
\setlength{\unitlength}{0.240900pt}
\begin{picture}(1049,900)(0,0)
\thicklines \path(985,113)(965,113)
\put(198,113){\makebox(0,0)[r]{$-0.15$}}
\thicklines \path(220,304)(240,304)
\thicklines \path(985,304)(965,304)
\put(198,304){\makebox(0,0)[r]{$-0.10$}}
\thicklines \path(220,495)(240,495)
\thicklines \path(985,495)(965,495)
\put(198,495){\makebox(0,0)[r]{$-0.05$}}
\thicklines \path(220,686)(240,686)
\thicklines \path(985,686)(965,686)
\put(198,686){\makebox(0,0)[r]{$0$}}
\thicklines \path(220,877)(240,877)
\thicklines \path(985,877)(965,877)
\put(198,877){\makebox(0,0)[r]{$0.05$}}
\thicklines \path(220,113)(220,133)
\thicklines \path(220,877)(220,857)
\put(220,68){\makebox(0,0){$0.5$}}
\thicklines \path(297,113)(297,133)
\thicklines \path(297,877)(297,857)
\thicklines \path(373,113)(373,133)
\thicklines \path(373,877)(373,857)
\put(373,68){\makebox(0,0){$0.6$}}
\thicklines \path(450,113)(450,133)
\thicklines \path(450,877)(450,857)
\thicklines \path(526,113)(526,133)
\thicklines \path(526,877)(526,857)
\put(526,68){\makebox(0,0){$0.7$}}
\thicklines \path(603,113)(603,133)
\thicklines \path(603,877)(603,857)
\thicklines \path(679,113)(679,133)
\thicklines \path(679,877)(679,857)
\put(679,68){\makebox(0,0){$0.8$}}
\thicklines \path(756,113)(756,133)
\thicklines \path(756,877)(756,857)
\thicklines \path(832,113)(832,133)
\thicklines \path(832,877)(832,857)
\put(832,68){\makebox(0,0){$0.9$}}
\thicklines \path(909,113)(909,133)
\thicklines \path(909,877)(909,857)
\thicklines \path(985,113)(985,133)
\thicklines \path(985,877)(985,857)
\put(985,68){\makebox(0,0){$1.0$}}
\thicklines \path(220,113)(985,113)(985,877)(220,877)(220,113)
\put(0,495){\makebox(0,0)[l]{\shortstack{$\bar{\cal A}_{\rm LP}$}}}
\put(602,23){\makebox(0,0){$\hat{s}$}}
\thinlines \path(295,623)(295,623)(296,596)(299,558)(302,529)(307,485)(312,451)(323,396)(333,354)(354,291)(375,246)(396,214)(417,190)(428,181)(438,173)(449,167)(459,163)(465,161)(470,159)(475,158)(480,157)(483,157)(486,156)(488,156)(490,156)(491,156)(492,
156)(493,156)(495,156)(496,156)(497,156)(499,156)(500,156)(501,156)(503,156)(504,156)(507,156)(509,156)(512,157)(517,157)(522,158)(533,161)(543,164)(564,173)(585,185)(606,200)(649,235)(691,279)(733,330)(775,389)(817,456)
\thinlines \path(817,456)(859,537)(880,589)(885,605)(888,615)(890,625)(893,638)(894,645)(896,654)(897,667)
\thicklines \path(294,659)(294,659)(294,643)(295,623)(296,608)(296,595)(299,557)(302,529)(307,487)(313,457)(316,445)(319,435)(321,427)(323,424)(324,421)(325,419)(326,419)(326,418)(327,418)(327,418)(328,418)(328,418)(328,418)(329,418)(329,418)(329,418)(330,418)(330,418)(330,418)(331,419)(332,420)(332,421)(334,424)(335,428)(337,433)(338,439)(341,457)(343,481)(346,512)(357,685)
\thicklines \path(357,682)(357,682)(358,710)(359,721)(359,731)(359,735)(359,739)(360,742)(360,744)(360,745)(360,746)(360,746)(360,747)(360,747)(360,747)(360,747)(360,747)(360,747)(361,747)(361,747)(361,747)(361,747)(361,747)(361,746)(361,746)(361,745)(361,743)(362,740)(362,732)(362,723)(363,699)(364,669)(366,603)(368,539)(369,483)(371,436)(373,398)(375,367)(376,342)(378,320)(380,302)(381,287)(383,274)(385,263)(387,254)(388,250)(388,247)(389,246)(389,245)(390,244)(390,243)
\thicklines \path(390,243)(390,243)(390,243)(390,243)(390,243)(391,243)(391,243)(391,243)(391,243)(391,243)(391,243)(391,243)(391,243)(391,243)(391,243)(391,243)(391,243)(391,243)(391,243)(392,243)(392,243)(392,244)(392,244)(393,245)(393,247)(394,249)(394,255)(395,262)(397,282)(399,308)
\thicklines \path(399,263)(399,263)(400,266)(400,268)(400,269)(401,270)(401,270)(401,271)(401,271)(401,271)(402,271)(402,272)(402,272)(402,272)(402,272)(402,271)(403,271)(403,271)(403,270)(404,269)(405,267)(410,245)(416,228)(422,215)(428,204)(434,196)(440,189)(445,182)(451,177)(457,173)(463,169)(469,165)(475,163)(480,160)(483,160)(486,159)(488,159)(489,158)(490,158)(491,158)(491,158)(492,158)(492,158)(493,158)(493,158)(493,158)(493,158)(493,158)(494,158)(494,158)(494,158)
\thicklines \path(494,158)(494,158)(494,158)(495,158)(495,158)(495,158)(495,158)(495,158)(496,158)(497,158)(497,158)(498,158)(499,159)(501,159)(502,159)(504,160)(507,162)(508,163)(510,164)(513,167)(516,171)(518,177)(521,185)(524,195)(527,208)(533,241)(539,286)
\thicklines \path(539,203)(539,203)(540,206)(541,208)(542,210)(544,211)(544,212)(545,212)(545,213)(546,213)(546,213)(547,213)(548,213)(548,213)(549,213)(549,213)(550,213)(550,213)(553,211)(557,208)(560,207)(562,206)(564,205)(567,204)(568,204)(569,203)(570,203)(571,203)(571,203)(572,203)(573,203)(573,203)(574,203)(574,203)(575,203)(575,203)(576,203)(577,203)(577,203)(578,203)(580,203)(581,203)(583,204)(585,204)(590,206)(595,207)(613,217)(632,230)(650,245)(669,262)(688,282)
\thicklines \path(688,282)(706,302)(725,324)(743,348)(762,374)(781,400)(799,429)(818,460)(836,493)(855,529)(864,549)(873,572)(883,598)(887,614)(890,623)(892,633)(894,646)(896,654)(896,658)(897,664)(897,672)
\end{picture}
        \end{center}
        \end{minipage}
        \hfill
        \begin{minipage}[t]{\minitwocolumn}
        \begin{center}
\setlength{\unitlength}{0.240900pt}
\begin{picture}(1049,900)(0,0)
\thicklines \path(220,113)(240,113)
\thicklines \path(985,113)(965,113)
\put(198,113){\makebox(0,0)[r]{$-0.8$}}
\thicklines \path(220,266)(240,266)
\thicklines \path(985,266)(965,266)
\put(198,266){\makebox(0,0)[r]{$-0.6$}}
\thicklines \path(220,419)(240,419)
\thicklines \path(985,419)(965,419)
\put(198,419){\makebox(0,0)[r]{$-0.4$}}
\thicklines \path(220,571)(240,571)
\thicklines \path(985,571)(965,571)
\put(198,571){\makebox(0,0)[r]{$-0.2$}}
\thicklines \path(220,724)(240,724)
\thicklines \path(985,724)(965,724)
\put(198,724){\makebox(0,0)[r]{$0$}}
\thicklines \path(220,877)(240,877)
\thicklines \path(985,877)(965,877)
\put(198,877){\makebox(0,0)[r]{$0.2$}}
\thicklines \path(220,113)(220,133)
\thicklines \path(220,877)(220,857)
\put(220,68){\makebox(0,0){$0$}}
\thicklines \path(373,113)(373,133)
\thicklines \path(373,877)(373,857)
\put(373,68){\makebox(0,0){$0.2$}}
\thicklines \path(526,113)(526,133)
\thicklines \path(526,877)(526,857)
\put(526,68){\makebox(0,0){$0.4$}}
\thicklines \path(679,113)(679,133)
\thicklines \path(679,877)(679,857)
\put(679,68){\makebox(0,0){$0.6$}}
\thicklines \path(832,113)(832,133)
\thicklines \path(832,877)(832,857)
\put(832,68){\makebox(0,0){$0.8$}}
\thicklines \path(985,113)(985,133)
\thicklines \path(985,877)(985,857)
\put(985,68){\makebox(0,0){$1.0$}}
\thicklines \path(220,113)(985,113)(985,877)(220,877)(220,113)
\put(602,23){\makebox(0,0){$\hat{s}$}}
\thinlines \path(220,724)(220,724)(225,623)(228,584)(231,551)(236,498)(242,455)(247,422)(253,394)(263,352)(274,323)(285,301)(296,286)(307,275)(312,270)(318,267)(323,263)(328,261)(334,259)(339,257)(345,256)(347,256)(350,255)(353,255)(354,255)(356,255)(357,255)(358,255)(358,255)(359,255)(360,255)(360,255)(360,255)(361,255)(361,255)(361,255)(362,255)(362,255)(362,255)(363,255)(363,255)(363,255)(364,255)(364,255)(365,255)(366,255)(368,255)(369,255)(372,255)(375,255)(377,256)
\thinlines \path(377,256)(383,256)(394,259)(404,261)(415,265)(426,268)(437,272)(448,277)(459,281)(464,283)(469,285)(472,286)(474,286)(475,287)(476,287)(476,287)(477,287)(477,287)(477,287)(478,287)(478,287)(478,287)(479,287)(479,287)(479,287)(480,287)(480,286)(480,285)
\thinlines \path(480,285)(480,285)(481,283)(482,282)(482,281)(483,281)(484,281)(484,281)(485,281)(486,281)(486,281)(487,281)(488,281)(490,282)(491,282)(496,284)(501,287)(522,301)(543,318)(564,335)(585,353)(606,372)(628,390)(649,410)(670,429)(691,448)(712,468)(733,488)(754,508)(775,528)(796,548)(817,568)(838,589)(859,610)(880,632)(901,654)(922,680)(932,695)(935,700)(936,703)(938,706)(939,710)(940,712)(940,715)(941,719)
\thicklines \path(220,724)(220,724)(221,705)(222,688)(222,672)(223,657)(224,643)(225,630)(226,618)(227,607)(227,597)(228,589)(229,581)(230,574)(230,572)(231,570)(231,569)(231,568)(231,568)(231,568)(231,567)(231,567)(232,567)(232,567)(232,567)(232,567)(232,567)(232,567)(232,567)(232,567)(232,567)(232,567)(232,567)(232,567)(232,567)(232,567)(232,567)(232,567)(232,567)(232,567)(232,568)(232,568)(233,569)(233,570)(233,572)(233,575)(234,580)(234,587)(235,596)(235,607)(236,620)
\thicklines \path(236,620)(236,657)(237,707)(238,769)(239,838)(239,877)
\thicklines \path(240,845)(240,845)(240,848)(240,851)(240,854)(240,858)(240,862)(240,867)(240,873)(240,877)
\thicklines \path(242,877)(243,775)(244,691)(245,626)(247,578)(248,542)(249,514)(250,491)(253,457)(255,431)(260,395)(265,369)(270,350)(280,320)(290,300)(300,284)(310,273)(320,265)(325,262)(330,259)(335,257)(340,255)(345,253)(350,252)(355,251)(358,251)(360,251)(363,250)(364,250)(365,250)(367,250)(368,250)(369,250)(370,250)(370,250)(371,250)(371,250)(372,250)(372,250)(372,250)(372,250)(373,250)(373,250)(373,250)(374,250)(374,250)(375,250)(375,250)(377,250)(378,250)(380,250)
\thicklines \path(380,250)(383,250)(385,250)(390,251)(400,252)(410,254)(420,256)(430,259)(440,261)(450,264)(460,267)(470,270)(480,272)
\thicklines \path(480,272)(480,272)(480,271)(480,271)(481,270)(481,270)(481,270)(481,269)(482,269)(482,269)(482,269)(482,268)(483,268)(483,268)(483,268)(484,268)(484,268)(484,268)(485,268)(485,268)(485,268)(485,268)(485,268)(485,268)(485,268)(485,268)(485,268)(486,268)(486,268)(486,268)(486,268)(486,268)(486,268)(486,268)(486,268)(486,268)(487,268)(487,268)(488,268)(488,268)(489,268)(490,268)(492,269)(495,270)(497,272)(500,273)(502,275)(505,278)(507,281)(509,284)(512,289)
\thicklines \path(512,289)(514,295)(517,303)(519,314)(521,329)(524,350)(526,380)(529,422)(531,480)(534,555)(536,641)(538,724)
\thicklines \path(538,724)(538,724)(539,758)(540,788)(540,800)(541,810)(541,814)(541,818)(541,820)(541,823)(541,824)(542,825)(542,826)(542,826)(542,825)(542,824)(542,822)(543,819)(543,813)(543,805)(544,784)(547,676)(548,625)(549,581)(551,544)(552,514)(554,490)(555,470)(556,454)(558,441)(559,430)(561,421)(563,407)(566,397)(569,389)(572,384)(574,380)(576,379)(577,377)(579,376)(580,376)(581,375)(583,374)(584,374)(585,373)(586,373)(587,373)(588,373)(588,373)(588,373)(589,373)
\thicklines \path(589,373)(589,373)(589,373)(589,373)(590,373)(590,373)(590,373)(590,373)(590,373)(590,373)(591,373)(591,373)(591,373)(591,373)(592,373)(592,373)(593,373)(594,373)(595,373)(596,374)(599,374)(605,376)(610,378)(616,380)(621,382)(627,385)(632,388)(638,391)(641,393)(643,395)(646,397)(649,401)(650,403)(652,406)(653,409)(655,413)(656,418)(657,424)(659,432)(660,443)(661,457)(663,476)(664,501)(666,533)(668,622)(671,723)
\thicklines \path(671,720)(671,720)(672,750)(672,762)(672,772)(672,776)(672,780)(672,782)(672,783)(672,784)(672,785)(672,785)(673,785)(673,785)(673,786)(673,786)(673,786)(673,786)(673,786)(673,786)(673,785)(673,785)(673,785)(673,785)(673,784)(673,783)(673,781)(673,775)(673,767)(674,757)(675,710)(675,662)(676,620)(677,586)(678,559)(679,538)(680,522)(681,508)(682,497)(682,487)(683,479)(684,472)(685,465)(686,459)(687,453)(688,448)(688,446)(688,445)(688,445)(689,444)(689,444)
\thicklines \path(689,444)(689,443)(689,443)(689,443)(689,443)(689,443)(689,443)(689,443)(689,443)(689,443)(689,443)(689,443)(689,443)(689,443)(689,443)(689,443)(689,443)(689,443)(689,443)(690,443)(690,444)(690,444)(690,445)(690,446)(690,448)(691,450)(691,455)(691,462)(692,470)
\thicklines \path(692,468)(692,468)(693,475)(693,478)(693,481)(694,483)(694,485)(694,485)(694,486)(695,486)(695,487)(695,487)(695,487)(695,487)(695,487)(695,487)(696,488)(696,488)(696,488)(696,488)(696,488)(696,488)(696,488)(696,488)(696,488)(696,487)(697,487)(697,487)(698,486)(699,485)(701,484)(702,483)(703,483)(704,482)(704,482)(705,482)(706,482)(706,482)(706,482)(707,482)(707,482)(707,482)(708,482)(708,482)(708,481)(708,481)(708,481)(708,481)(708,481)(708,481)(708,481)
\thicklines \path(708,481)(709,481)(709,481)(709,481)(709,481)(709,481)(709,481)(709,481)(709,482)(710,482)(710,482)(711,482)(712,482)(712,482)(715,482)(718,483)(721,484)(724,485)(727,487)(730,488)(733,489)(736,490)(739,491)(741,492)(744,493)(747,493)(749,494)(750,495)(752,495)(752,496)(753,497)(754,498)(755,499)(755,500)(756,502)(758,506)(759,512)(760,520)(762,531)
\thicklines \path(762,516)(762,516)(764,524)(765,528)(767,531)(768,533)(769,535)(770,537)(771,538)(776,542)(781,545)(790,551)(799,558)(808,566)(818,574)(827,583)(836,591)(846,600)(855,609)(864,618)(873,627)(883,636)(892,646)(901,656)(911,666)(920,678)(929,691)(934,698)(936,703)(937,706)(939,709)(940,712)(940,715)(941,716)(941,718)(941,720)
\thinlines \dashline[-10]{25}(220,724)(225,606)(228,562)(231,524)(236,463)(242,416)(247,379)(253,349)(258,325)(263,305)(269,288)(274,275)(280,263)(285,254)(291,245)(296,239)(307,229)(312,225)(318,222)(323,219)(328,218)(331,217)(334,216)(337,216)(339,215)(341,215)(342,215)(343,215)(344,215)(345,215)(345,215)(346,215)(346,215)(347,215)(347,215)(347,215)(348,215)(348,215)(348,215)(349,215)(349,215)(349,215)(350,215)(350,215)(351,215)(352,215)(353,215)(354,215)(356,215)(358,215)
\thinlines \dashline[-10]{25}(361,216)(366,217)(372,218)(383,221)(394,224)(404,228)(415,233)(426,238)(437,243)(448,249)(459,255)(464,258)(469,260)(472,261)(475,263)(476,263)(477,263)(477,263)(478,263)(478,263)(478,263)(479,264)(479,264)(479,263)(480,263)(480,263)(480,262)
\thinlines \dashline[-10]{25}(480,262)(501,276)(522,294)(543,313)(564,331)(585,350)(606,370)(628,389)(649,409)(670,428)(691,448)(712,468)(733,487)(754,507)(775,528)(796,548)(817,568)(838,589)(859,610)(880,632)(901,654)(922,680)(932,696)(935,700)(936,703)(938,706)(939,710)(940,712)(940,715)(941,719)
\thinlines \dashline[-10]{25}(220,724)(225,583)(228,531)(231,488)(236,418)(242,366)(247,326)(253,294)(258,269)(263,248)(269,232)(274,218)(280,207)(285,198)(291,191)(296,185)(301,180)(307,176)(312,174)(315,173)(318,172)(320,171)(323,170)(326,170)(327,170)(328,169)(330,169)(330,169)(331,169)(332,169)(332,169)(333,169)(333,169)(333,169)(334,169)(334,169)(334,169)(335,169)(335,169)(335,169)(336,169)(336,169)(337,169)(337,169)(338,169)(339,169)(341,170)(342,170)(345,170)(350,171)
\thinlines \dashline[-10]{25}(356,172)(361,174)(372,178)(383,182)(394,188)(404,194)(415,200)(426,207)(437,214)(448,222)(459,229)(469,237)(480,243)
\thinlines \dashline[-10]{25}(480,243)(501,267)(522,288)(543,309)(564,329)(585,349)(606,368)(628,388)(649,408)(670,428)(691,448)(712,468)(733,488)(754,508)(775,528)(796,548)(817,568)(838,589)(859,610)(880,632)(901,654)(922,680)(932,696)(935,700)(936,703)(938,706)(939,710)(940,712)(940,715)(941,719)
\thinlines \dashline[-10]{25}(220,724)(223,611)(224,571)(225,536)(228,479)(231,435)(234,398)(236,368)(242,322)(247,289)(253,263)(258,244)(263,230)(269,219)(274,210)(280,204)(285,199)(288,197)(291,195)(293,194)(296,193)(299,192)(300,192)(301,192)(303,191)(304,191)(305,191)(306,191)(307,191)(307,191)(307,191)(308,191)(308,191)(308,191)(309,191)(309,191)(309,191)(310,191)(310,191)(311,191)(311,191)(312,191)(312,191)(314,191)(315,191)(318,192)(320,192)(323,193)(328,194)(339,198)
\thinlines \dashline[-10]{25}(350,204)(361,210)(372,217)(383,224)(394,232)(404,241)(415,250)(426,259)(437,269)(448,280)(459,291)(464,297)(469,304)(472,307)(475,311)(476,314)(478,316)(479,320)(479,321)(480,322)(480,323)(480,326)
\thinlines \dashline[-10]{25}(480,326)(481,325)(482,325)(482,324)(483,324)(484,324)(484,324)(485,324)(486,324)(486,324)(487,324)(488,324)(490,325)(491,325)(496,326)(501,328)(522,338)(543,349)(564,363)(585,378)(606,393)(628,409)(649,426)(670,443)(691,461)(712,479)(733,497)(754,516)(775,534)(796,554)(817,573)(838,593)(859,613)(880,634)(901,656)(922,681)(932,696)(935,701)(936,704)(938,707)(939,710)(940,712)(940,715)(941,719)
\end{picture}
        \end{center}
        \end{minipage}
        \caption{Left : LP asymmetry for $\tau^+ \, \tau^-$ 
	with (solid thick curve) and without (solid thin curve) 
	resonances for $(\gamma, x) = ({48.6}^o, 0.778)$. 
	Right : same with the left but for $\mu^+ \, \mu^-$.
	The dashed curves show the SD contribution for 
	same $x$ and $\gamma = 0^o, {90}^o, {-90}^o$.}
        \label{fig:lpa}
\end{figure}
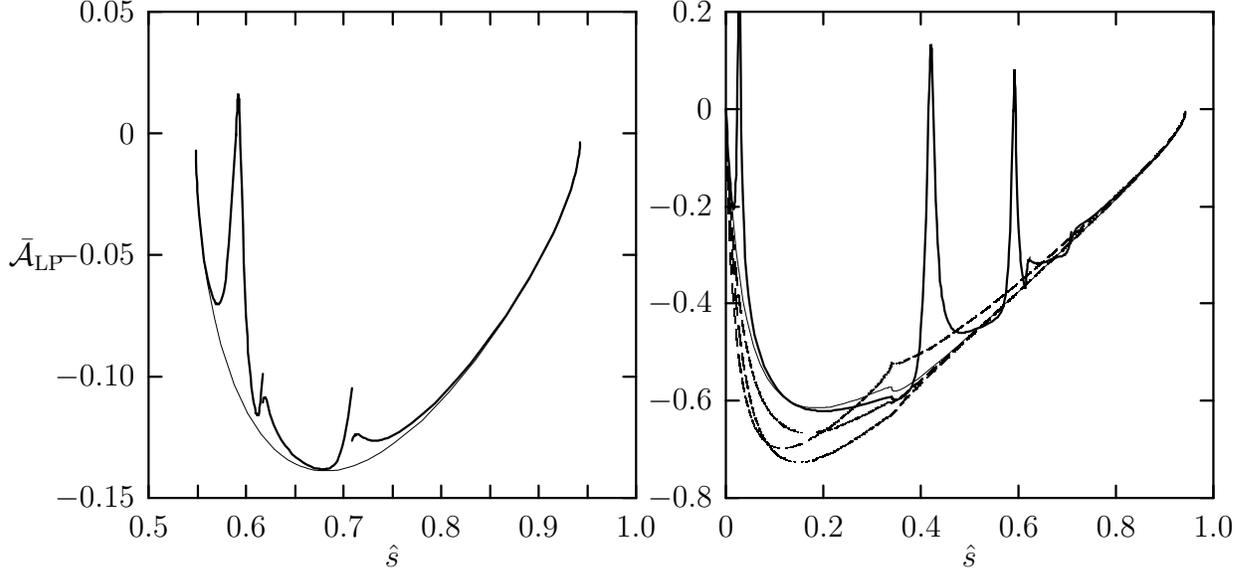

Until now, all of the measurements are less
sensitive to the lepton mass. Next, I provide the LP asymmetry 
which must be considered for heavy dilepton final state and is 
proposed first in \cite{lpa} for $B \rightarrow X_s \, \tau^+ \, \tau^-$.
Generally, the normalized LP asymmetry is given as
\begin{equation}
	\bar{\cal A}_{\rm LP}(\s) = 
	\frac{\displaystyle
	{{\rm d}{\cal B}(\s,\mbox{\boldmath $n$})}/{{\rm d}\s}
	- 
	{{\rm d}{\cal B}(\s,-\mbox{\boldmath $n$})}/{{\rm d}\s}
	}{\displaystyle 
	{{\rm d}{\cal B}(\s,\mbox{\boldmath $n$})}/{{\rm d}\s}
	+ 
	{{\rm d}{\cal B}(\s,-\mbox{\boldmath $n$})}/{{\rm d}\s}
	}	
	= \frac{{{\rm d}{\cal A}_{\rm LP}(\s)}/{{\rm d}\s}
	}{{{\rm d}{\cal B}(\s)}/{{\rm d}\s}} \; ,
	\label{eq:nlpa}
\end{equation}
with $\mbox{\boldmath $n$}$ is a unit vector of any given spin direction 
of $\l^-$ in its rest frame. Then, for the longitudinal polarization,
that is $\mbox{\boldmath $n$}$ has same direction with the momentum  
of $\ell^-$ ($\mbox{\boldmath $p$}_{\ell^-}$), 
\begin{eqnarray}
	\frac{{\rm d}{\cal A}_{\rm LP}(\s)}{{\rm d}\s} 
	& = & \frac{8}{3} \, {\cal B}_o \, 
	\left(1 - \frac{4 \,\ml^2}{\s} \right) \, \u(\s) \, C_{10} \, 
	\left\{ 6 \, {C_7}^{\rm eff} \, \left[ (1 - \mq^2)^2 - 
		\s \, (1 + \mq^2) \right] 
	\right.
	\nonumber \\
	& & \left. \; \; \; \; \; \; 
	+ {\rm Re} \left( {C_9}^{\rm eff} \right)^\ast \, 
	\left[ (1 - \mq^2)^2 + \s \, (1 + mq^2) - 2 \, \s^2 \right] 
	\right\} \; .
	\label{eq:dlpa}
\end{eqnarray}

The distribution is depicted in Fig. \ref{fig:lpa}. 
As shown in the figure, for $\tau^+ \tau^-$ final state the sensitivity on 
$\gamma$ is tiny. The reason is, for $\tau^+ \tau^-$ final state the 
distribution starts appearing in the region higher than $\s
= (4 \, {\hat{m}_\tau}^2)$, 
on the other hand, generally the high sensitivity on $\gamma$ is expected 
in the low $\s$ region. 
So it may be interesting to consider the transversal polarization  
that has different structure \cite{lpa}. 
Unfortunately, I have checked that the transversal lepton polarization 
asymmetry is too small for light dilepton like $\mu^+ \mu^-$, 
then one must consider $\tau^+ \tau^-$ again that the distribution 
is limited for higher $\s$ region.

\section{\bf Summary}
\label{sec:discussion}

I have shown how to extract the angle $\gamma$ of CKM unitarity 
triangle by the inclusive $B \rightarrow X_d \, \l^+ \, \l^-$ 
decay and the data of $x_d$ in the SM. 
As the results, finally I can make some points as below.
\begin{enumerate}
\item For low $\s$ region, i.e. $0.1 < \s < 0.3$, $5\sim10\%$
	discrepancies in ${\cal B}(B \rightarrow X_d \, e^+ \, e^-)$
	and $\bar{\cal A}_{\rm LP}(B \rightarrow X_d \, \mu^+ \, \mu^-)$
	may be good signals for the CP violation factor
	defined here. 
\item According to the fact that the sensitivity on $\gamma$ in the high $\s$ 
	region, i.e. $\s > 0.6$, is tiny, a measurement of $x$ 
	may be done by exploring one of the measurements discussed 
	in the present paper in addition to the present data of 
	$x_d$ in $B_d^0-\bar{B}_d^0$ mixing.
\item CP violation asymmetry in the channel should measure the dependence 
	on $\gamma$. It will, at least, be a good probe to determine 
	the sign of the angle $\gamma$.
\end{enumerate} 
  
To conclude, the measurements of the decay rate and 
asymmetries in $B \rightarrow X_d \, \l^+ \, \l^-$ decay 
will provide an independent information for $\gamma$. 
The information is a crucial test of CKM unitarity as well as 
leading to the discovery of unitarity violation. 

\bigskip
\noindent
{\Large \bf Acknowledgements}

I thank T. Muta for reading the manuscript, 
M. R. Ahmady and Y. Kiyo for useful discussion in $q-$ dependence in 
the long distance effects and D. X. Zhang for pointing 
out the $1/{q^2}$ dependence in the leptonic decay of vector mesons.
I also would like to thank Particle Elementer Physics Group for the
warm hospitality during my stay at ICTP Italy in the last part of the 
work and the Ministry of Education, Science and Culture (Monbusho-Japan) 
for the financial support.

\end{document}